\documentclass[11pt]{article}
\usepackage[english]{babel}
\usepackage{amsmath} 
\usepackage{amssymb}
\usepackage{amsfonts}
\usepackage{amsthm}
\usepackage[table]{xcolor}
\usepackage{tikz}
\usepackage{enumerate}
\usepackage{graphicx}
\usepackage[utf8]{inputenc}

\usepackage{hyperref}
\usepackage{url}
\usepackage{epstopdf} 
\usepackage{geometry}
\usepackage{cite}
\usepackage[nottoc]{tocbibind}
\usepackage{tikz}
\usetikzlibrary{shapes, shadows, arrows}
\tikzstyle{block}=[draw,rectangle,fill=blue!5,text width=12 em,text centered, minimum height=12mm, node distance=5 em]
\tikzstyle{line} = [draw,-latex']
\begin{document}
	\title{\textbf{Multi-point Infection Dynamics of Hepatitis B in the Presence of Sub-Viral Particles}}
	\date{}
\author{Rupchand Sutradhar, Gopinath Sadhu and   D C Dalal\\Indian Institute of Technology Guwahati, Guwahati, Assam, India, 781039}
	\maketitle
	\noindent
	
	\section{Abstract}
	\noindent
	Hepatitis B virus (HBV) is considered as  etiological agent of the lethal liver disease hepatitis B.
	Globally, hepatitis B is recognized as one of the prevailing infectious diseases with a significant impact on human health. In spite of being non-infectious in nature,  sub-viral particles (SVPs) , composed with mainly viral surface proteins, play  critical roles in the persistence and progression of the infection. Although the understanding on the functions of these non-infectious SVPs remains limited and incomplete. 
	In this study, a mathematical model is proposed for the first time  by incorporating the roles of SVPs and including the effects of capsids recycling. The impacts of spatial mobility of capsids, viruses, SVPs and antibodies are also taken into account in this model. Overall, this model carry unique
	characteristics in the context of this viral infection. This study investigates the changes in the dynamics of infection considering both single-point as well as multi-point infection initial condition. As a result, it is observed that  SVPs can significantly enhance intracellular viral replication and gene expression by  reducing the
	neutralization of virus particles by antibodies. The recycling of capsids substantially increase the concentration of SVPs. The experiments which are carried out on single-point as well as multi-point infection initial conditions show that if the liver is
	infected at more than one point, infection propagates rapidly. The outcomes of the proposed model strongly recommend that it is imperative to consider the diffusion term while HBV infection dynamics are illustrated. \vspace{0.5cm}\\
	Keywords: Hepatitis B; Sub-viral particles; Mathematical model; Diffusion; Multi-point infection;  Numerical simulation
	\section{Introduction}
	\noindent
	Hepatitis B virus (HBV) represents a significant and formidable threat to human health throughout the world  with  ample potential to cause life-threatening consequences. This viral infection infiltrates the liver, leading to a wide spectrum of outcomes that can extend from acute illness to chronic infection, and in some cases, even to fatal complications like hepatocellular carcinoma (HCC) \cite{WHO_2021}. The insidious nature of HBV lies in its ability to silently persist within the host body, quietly replicating and damaging liver cells over an extended period of time. Left untreated or undiagnosed, chronic HBV infection can progress to severe conditions such as liver cirrhosis where the liver becomes scarred and loses its functionality, ultimately resulting liver  failure. The devastating impacts of HBV on individuals and communities emphasize the urgent need for widespread awareness, vaccination, early diagnosis, and access to effective treatment interventions to mitigate this life-threatening consequences.

	Over the past two decades, many authors have been diligently researching HBV infections from different angles \cite{2008_Min,2016_jun_nakabayashi,2018_fatehi_nkcell,2015_Murray,2021_ciupe_early}.  The primary causes for the persistence of this infection include the high stability of cccDNAs, the intricate life cycle of the virus, and the crucial functions played by certain viral components, such as HBx proteins, surface proteins (L/M/SHBs), sub-viral particles (SVPs), and many more.  SVPs which are produced by HBV infected cells during infection period are  non-infectious in nature.
	%
	%
	These are predominantly composed of viral surface antigens (HBsAg)  proteins that are the major components of the  lipid bilayer envelope \cite{2021_prifti}. SVPs don't contain nucleocapsid proteins and viral nucleic acids, \textit{i.e.}, lack the rcDNA-containing capsids or the viral genome. 
	The production of SVPs is a natural biological process. While the viruses replicate within the hepatocytes (liver cells), both complete viral particles and SVPs  produce simultaneously and release into the bloodstream.  Generally, SVPs are found  in the blood of infected individuals  in two main forms: (i) spherical particles measuring 25 nm in diameter and (ii)  filaments with a diameter of 22 nm which can differ in length \cite{2005_gripon_efficient,2013_gerlich_medical}. 
	In the literature, many authors have reported that the ratio of Dane particles to SVPs ranges from 1:10,000 to 1:100,000 \cite{2013_gerlich_medical,2015_luckenbaugh_genome,2017_hu_complete} , \textit{i.e.}, a large quantity of HBsAg derived from viral particles has no effect on the reservoir of viral replication. Although,  the actual reasons behind their overproduction and their exact roles in HBV pathogenesis are still under investigation \cite{2009_garcia_drastic}. The assembly and secretion pathways for SVPs are typically distinct from those used by the virus.  
	Researchers are dedicated to unravel the primary causes underlying their excessive production and understand their critical functions in the development of HBV-related illnesses. In individuals with chronic HBV infection, the number of SVPs far surpasses that of infectious virions in the bloodstream. To the best of our knowledge, despite its significant overproduction, the functions of SVPs have been widely overlooked in the previous studies \cite{1996_Nowak,2005_murray_dynamics,2007_Ciupe_Role,2007_Dahari,2007_wang_wang,2009_Eikenberry,2010_Hews,2011_Jun_nakabayashi,2015_Murray,2018_Danane_mathematical,2018_Guo,2021_Fatehi,2021_ciupe_early,2023_sutradhar_fractional}, leaving a gap in our knowledge about their biological importance. Recently, Cao et al. \cite{2019_cao_cryo} demonstrated the distribution of Dane particles (virus), spherical particles, and filamentous particles using micrographs and
	suggested that there is  a universal folding shape of HBsAg on  SVPs.
	From the perspective of Rydell et al.\cite{2017_rydell_hepatitis,2019_cao_cryo}, SVPs are believed to potentially facilitate cell-to-cell spread, particularly in the presence of neutralizing antibodies.  SVPs function as a disguise for the virus. As a result of this characteristic of SVPs, a substantial portion of HBV-specific antibodies produced by the host's immune system is targeted towards neutralizing SVPs, thereby helping the virus evade the immune response \cite{2017_rydell_hepatitis}. However, to the best of our knowledge, this effect has not been demonstrated experimentally or in animal models so far. Although,  SVPs are less efficient in blocking viral entry by binding to target cells \cite{2008_chai_properties}.  High levels of SVPs can stimulate the immune system, leading to the production of antibodies against HBsAg.
	For this virus, Bruns et al. \cite{1998_bruns_enhancement},  and Klingmüller and Schaller \cite{1993_klingmuller_hepadnavirus} reported that SVPs have capability to both enhance and inhibit cell attachment of the virus. The collective evidence suggests that SVPs have some notable impacts  on the advancement and perpetuation of the infection. Based on these findings, it is apparent that more elaborate studies are imperative to enhance  understanding about SVPs.
	
	The propagation  of viral particles both between and within the hosts is inherently a spatial process. Diffusion plays a pivotal role in viral infection by facilitating the spread of viral particles both locally within the liver and systemically throughout the body. This diffusion also impacts the immune response, as SVPs diffuses widely, potentially diverting immune attention from infectious virions and contributing to immune evasion.
	Some researchers have already addressed how spatial heterogeneity  influences the dynamics of infection of  certain viruses, such as HIV \cite{2021_wu_dynamical},  Covid-19 \cite{2023_wu_spatial}, etc. In case of HBV infection, a number of studies \cite{2023_miao_dynamics,2009_xu_hbv,2008_wang_dynamics} have also been  documented  the effects of diffusion of virus particles in the literature. However, none of these studies have simultaneously taken into account the effects of capsid recycling, the roles of SVPs, and the functions of antibodies. Recently,  Sutradhar and Dalal \cite{2023_sutradhar_recycling,2023_sutradhar_fractional} have explored the impacts of capsid recycling on HBV infection and found that the cytoplasmic recycling of rcDNA-containing capsids enhances the infection. So, it is important to consider capsid recycling while studying HBV infection. 
	
	
	Motivated by the aforementioned discussion, in this study, we shed light on the roles of  SVPs in the development of infection via a modified HBV infection dynamics model.  In this framework,  incorporating the diffusion   of capsids, virus, SVPs and antibodies, and the impacts of SVPs,  we extend our previously  proposed HBV infection dynamics model mentioned in the article \cite{2023_sutradhar_recycling}. The precise interplay between virus particles and antibodies is also focused on this work.   In this immediate version, it is considered that the capsids, viruses, SVPs and antibodies follow the Fickian diffusion that  refers to the process of molecular or particle transport driven by concentration gradients, described by Fick's laws of diffusion. The influences of the spatial diffusion of capsids, viruses, SVPs, and antibodies will be also  explained from different perspectives.
	On the other hand, the initial conditions of the infection are another key factor that significantly affects the infection's dynamics. It is also aimed to investigate the correlations between outcomes of the infection and various types of initial conditions, such as single-origin, dual-origin, triple-origin, etc. Here,``$n$-origin" initial conditions indicate that the infection originates from $n$ distinct points within the liver.
	The following are the main contributions of this study:
	\begin{itemize}
		\item The roles of SVPs on HBV infection.
		\item The impacts spatial mobility of HBV DNA-containing capsids, viruses, SVPs and antibodies.
		\item The effects of capsid recycling  on the production of SVPs.
		\item Dysfunction of antibody in the presence of SVPs.
		\item Dual-origin and triple-origin infection: Unraveling the dynamics of infection.
	\end{itemize}
	\section{Formulation of model based on existing literature}
	
	\noindent
	In the last three decades, extensive research has been carried out  with the goal of eradicating HBV  infection and ensuring human well-being. Many researchers have proposed  different mathematical models to study the spread of infection. Each model was designed with a specific objective and pursued in a distinct direction \cite{2009_Eikenberry,2010_Hews,2011_Jun_nakabayashi,2015_Murray,2018_Danane_mathematical,2018_Guo,2021_Fatehi,2021_ciupe_early,2023_sutradhar_fractional}. For example,  Sutradhar and Dalal \cite{2023_sutradhar_recycling} recently studied the effects of capsid recycling in HBV infection by proposing the following HBV infection dynamics model (with slight symbolic modification):	
	\begin{align} \label{model_rup}
	\dfrac{dX(t)}{dt}&=\lambda-\mu X(t)-kV(t)X(t),\nonumber\\
	\dfrac{dY(t)}{dt}&=kV(t)X(t)-\delta Y(t),\nonumber\\
	\dfrac{dU(t)}{dt}&=aY(t)+\gamma(1-\eta)U(t)-\eta\beta U(t)-\delta U(t),\\
	\dfrac{dV(t)}{dt}&=\eta \beta U(t)-\delta_v V(t).\nonumber 
	\end{align}
	The temporal change of each component of this proposed model \eqref{model_rup} is determined based on the mass action principle. Here, the variables $X(t)$, $Y(t)$, $U(t)$, and $V(t)$ are ascribed to the specific compartment of the system of equations \eqref{model_rup}, representing uninfected hepatocytes, infected hepatocytes,  rcDNA-containing capsids, and  viruses, respectively. Uninfected hepatocytes are assumed to be produced with the natural growth rate of $\lambda$ and  degraded with natural death rate $\mu$. It is considered that the  infectious virus particles have the capability to infect the uninfected hepatocytes  at a rate of $k$. The parameter $a$ denotes the production rate of capsids  from infected hepatocytes. Within this model \eqref{model_rup},  $\eta$ fraction  of the capsids is assumed to give birth of new viruses with a production rate of $\beta$. On the other hand,  the remaining portion of  the capsids is returned to the nucleus at a rate $\gamma$ to increase the pool of covalently closed circular DNA (cccDNA). The process where newly produced capsids go back to the nucleus is termed  ``recycling of capsids", with $\gamma$ representing the recycling rate. Both infected hepatocytes and capsids undergo natural decay at a common death rate $\delta$. The parameter, $\delta_v$ denotes the spontaneous clearance rate of viruses. 
	
	In order to extend the above model \eqref{model_rup}, authors of this article incorporate the functions of SVPs  and antibodies through the introduction of two  distinct additional classes, namely SVPs and antibodies. It is believed that the diffusion of viral components plays a pivotal role in controlling and preventing infection.   The  effects of spatial mobility of all viral components (rcDNA-containing capsids, viruses, SVPs) and antibodies are also taken into account to make the model more reliable and closer to reality. 

	\subsection{Governing equations}
\begin{figure}
		\noindent
		\includegraphics[
		page=1,
		width=\textwidth,
		height=\textheight,
		keepaspectratio
		]{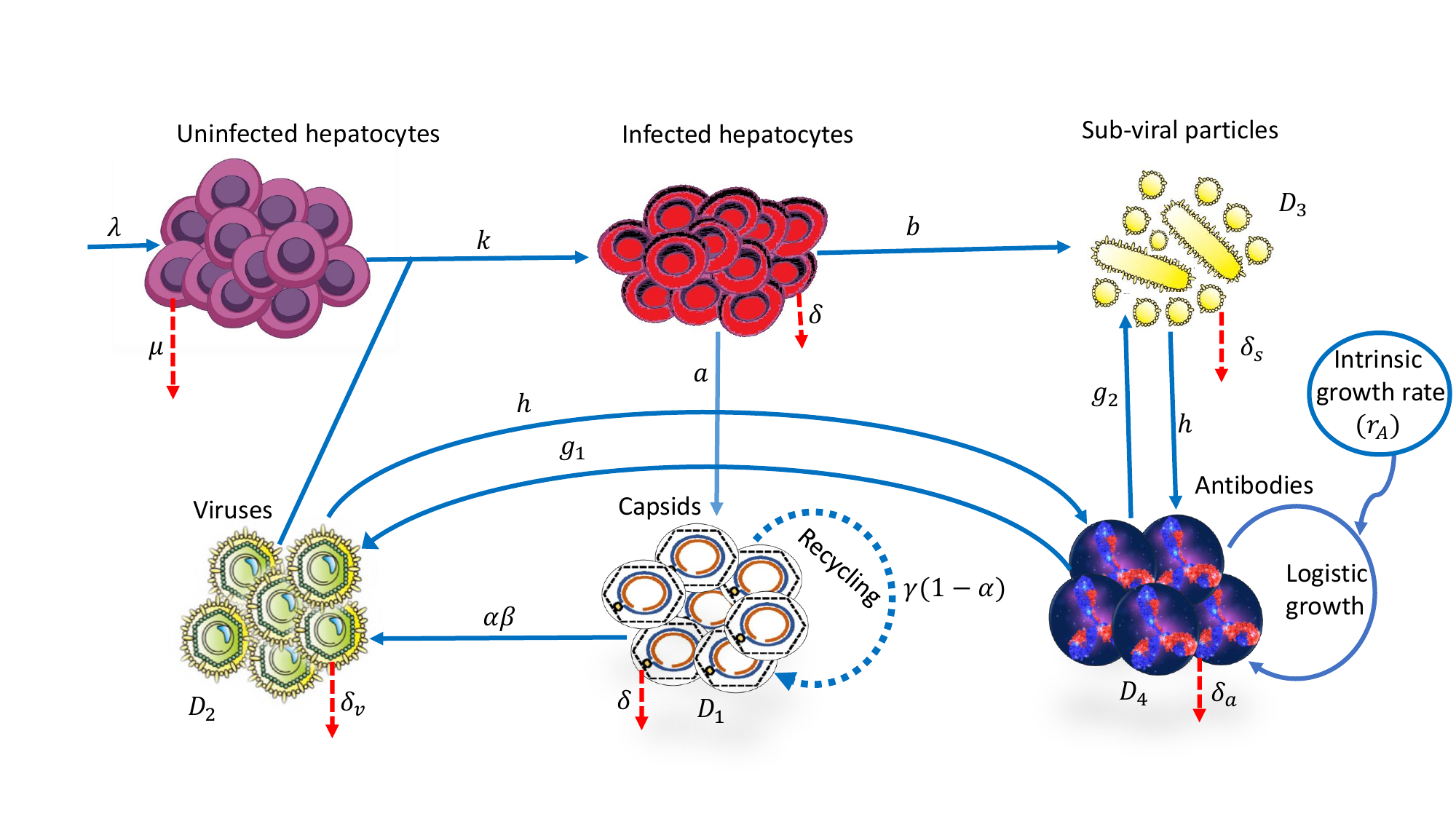}
		\vfill
		\newpage
		\caption{Semantic representation of the proposed model  given by the system of equations \eqref{eq:Healthy cells}-\eqref{eq:antibody}.}
		\label{Semantic representations}
	\end{figure}

	The mathematical formulation of the  model is as follows:
	\begin{align}
		\frac{\partial X(x,t)}{\partial t}&= \lambda-\mu X(x,t)-kV(x,t)X(x,t),~~\mbox{in}~~ \Omega\times (0,\tau),
		\label{eq:Healthy cells}\\
		\frac{ \partial Y(x,t)}{\partial t}&=kV(x,t)X(x,t)-\delta Y(x,t),~~\mbox{in}~~ \Omega\times (0,\tau),
		\label{eq:infected cells}\\
		\frac{\partial U(x,t)}{\partial t}&=\mathcal{D}_1\dfrac{\partial^2 U(x,t)}{\partial x^2}+aY(x,t)+\gamma(1-\alpha)U(x,t)-\alpha\beta U(x,t)-\delta U(x,t),~~\mbox{in}~~ \Omega \times (0,\tau),
		\label{eq:capsids}\\
		\frac{\partial V(x,t)}{ \partial t}&=\mathcal{D}_2\dfrac{\partial^2 V(x,t)}{\partial x^2}+\alpha\beta U(x,t)-g_1 A(x,t) V(x,t)-\delta_v V(x,t),~~\mbox{in}~~ \Omega\times (0,\tau),
		\label{eq:virus}\\
		\frac{\partial S_v(x,t)}{ \partial t}&=\mathcal{D}_3\dfrac{\partial^2 S_v(x,t)}{\partial x^2} +bY(x,t)-g_2 A(x,t) S_v(x,t)-\delta_{s} S_v(x,t),~~\mbox{in}~~ \Omega\times (0,\tau),
		\label{eq:subviral}\\
		\frac{\partial A(x,t)}{\partial t}&=\mathcal{D}_4\dfrac{\partial^2 A(x,t)}{\partial x^2}+h(V(x,t)+S_v(x,t))+r_AA(x,t)\left(1-\dfrac{A(x,t)}{A_m}\right)-g_1A(x,t)V(x,t)\nonumber\\&~~~~-g_2A(x,t)S_v(x,t)-\delta_a A(x,t),~~\mbox{in}~~ \Omega \times (0,\tau),
		\label{eq:antibody}
	\end{align}
	
\noindent
	where $\Omega\subset \mathbb{R}$  is a bounded domain where viruses, cells and antibodies  stay and can interact. Here,  $X(x,t)$, $Y(x,t)$, $U(x,t)$, $V(x,t)$, $S_v(x,t)$ and $A(x,t)$ represent uninfected cells, infected cells, rcDNA-containing capsids, viruses, SVPs, and antibodies at the point $(x,t)\in \Omega\times (0,\tau)$, respectively. The physical meaning of the parameters $\lambda, \mu, k, \delta, a, \gamma, \eta, \beta, \delta_v$ are the same as those in  model \eqref{model_rup}. We adhere to the fact that uninfected and infected hepatocytes do not exhibit any kind of spatial movement. The corresponding spatial mobilities of capsids, viruses, SVPs, and antibodies in the liver are characterized by diffusion coefficients $\mathcal{D}_1$, $\mathcal{D}_2$, $\mathcal{D}_3$, and $\mathcal{D}_4$. The proposed model (equations \eqref{eq:Healthy cells}-\eqref{eq:antibody}) will be called \textit{diffusion model} if  the value of any $\mathcal{D}_i,~i=1,2,3,4$ is non-zero. Otherwise, this model will be addressed as \textit{non-diffusion model}. The parameters $b$ denote the production rate of SVPs from infected hepatocytes. In general, $b>>a$ since infection with HBV results in the synthesis of a large number of SVPs, likely at least 1,000-fold more than virus particles \cite{2012_prange_host}. 
	It is considered that  the development of antibodies  is equally influenced by the presence of both virions and SVPs.  The term $h(V(x,t)+S_v(x,t))$ reflects the production of antibodies whereas $h$ denotes the production rate of antibodies following the work done by Ciupe et al. \cite{2014_ciupe_antibody}. Since both virions and SVPs display the HBV-specific antigen on their surface, antibodies are equally effective in neutralizing both types, thereby providing a robust immune response against infection. The neutralizing rate of viruses is given by  $g_1A(x,t)V(x,t)$ as described in the article of Chenar et al. \cite{2018_chenar_mathematical}, where $g_1$ represent the association rate constant for antibodies reacting to viral.The SVPs are also inactivated by direct binding with antibodies, which is represented by the term $g_2A(x,t)S_v(x,t)$ in the model \cite{2014_ciupe_antibody,2019_huang_mathematical}. The parameter $\delta_{s}$ and $\delta_a$  denote the normal  decay rates of  SVPs and antibodies, respectively. In the absence of the virus, we also introduce a logistic term with a maximum intrinsic growth rate  $r_A$ and a carrying capacity $A_m$ for the maintenance of antibodies through homeostasis. This term ensures that the antibody population grows to a stable level, reflecting the natural homeostatic control mechanisms within the immune system.In a nutshell,  the semantic representation of the above proposed model (equations \eqref{eq:Healthy cells}-\eqref{eq:antibody}) are shown in the Figure \ref{Semantic representations}. This model encompasses numerous biological phenomena and can be regarded as more generalized than existing models.
	\subsection{Initial and boundary conditions}
\begin{figure}[h]
		\noindent
		\includegraphics[
		page=1,
		width=\textwidth,
		height=\textheight,
		keepaspectratio
		]{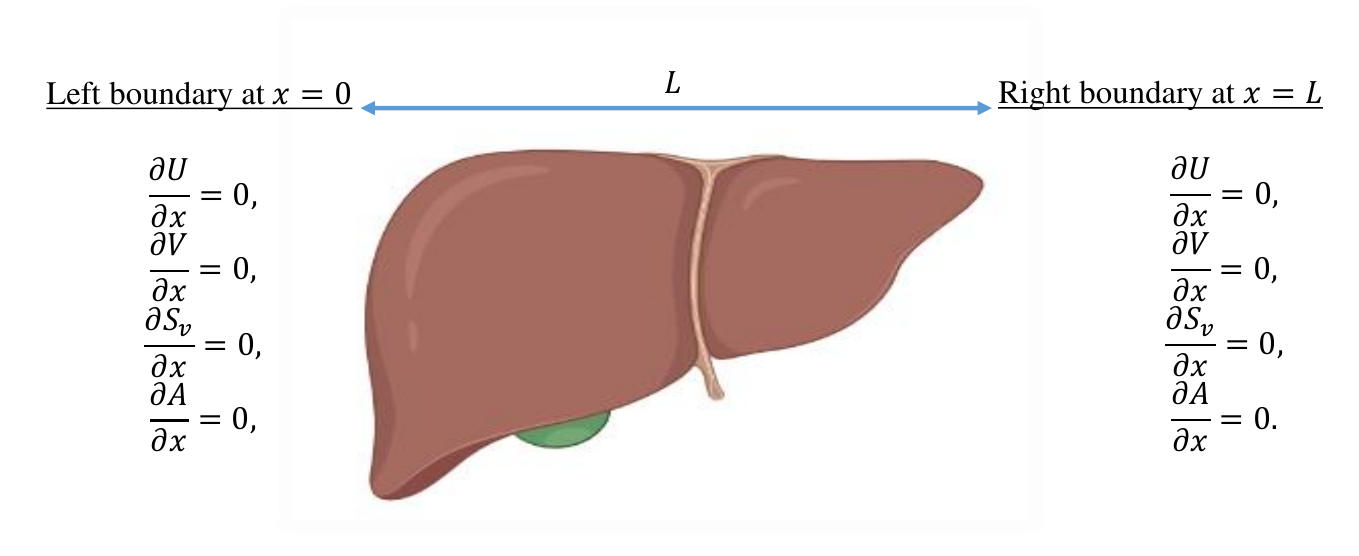}
		\vfill
		\caption{The length of the liver and the prescribed boundary conditions at $x=0$ and $x=L$ are shown at the left and right sides of the liver, respectively. This
			picture of liver is taken from BioRender.com.}
		\label{liver}
	\end{figure}

\begin{figure}[h!]
		\centering
		\includegraphics[width=15cm, height=10cm]{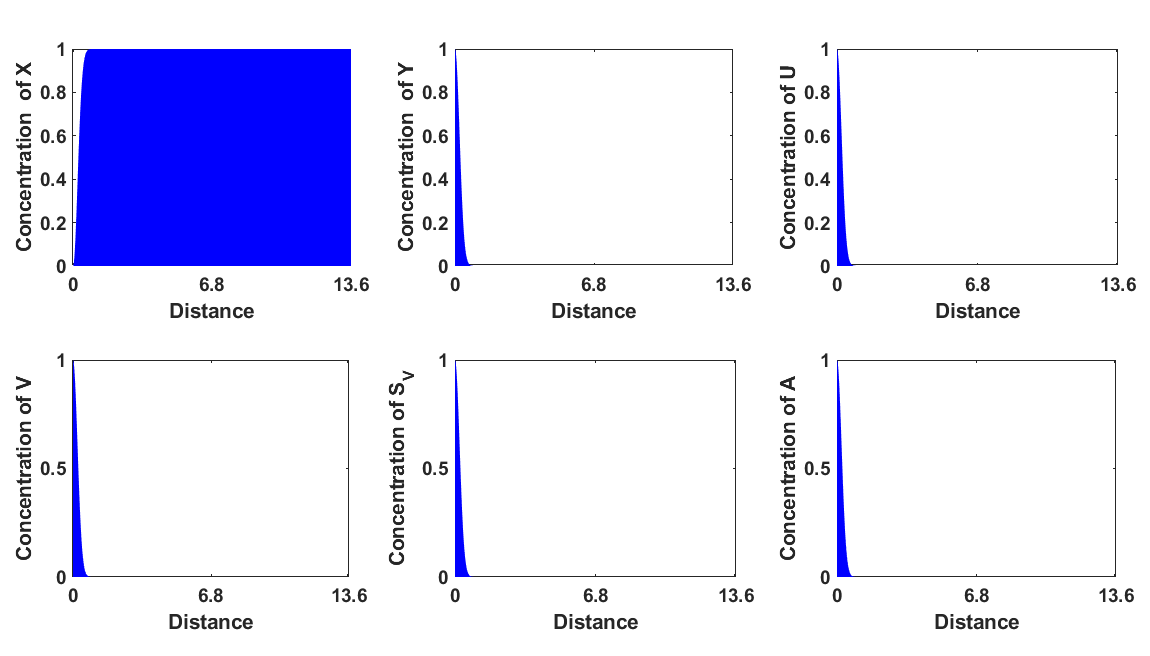}
		\caption{The graphical representation of all initial conditions mentioned in the equation \eqref{eq:initial}.}
		\label{initial_condition_all}
	\end{figure}

	\noindent
	In order to close the system (equation \eqref{eq:Healthy cells}-\eqref{eq:antibody}), it is essential to specify reasonable initial and boundary conditions for each compartment ($X(x,t), Y(x,t), U(x,t), V(x,t), S_v(x,t),$ and $A(x,t)$) of the proposed model.  For the convenience of analysis, let $L$ be the total length of the liver shown in Figure \ref{liver} and $[0,L]$ be the domain of interest. We denote $[0, L]$ as $\Omega$. It is assumed that the infection initially starts from $x=0$, \textit{i.e.}, from the left side of the liver. 
	\begin{itemize}
		\item \textbf{Initial conditions:} 
	\end{itemize} 
	For defining the  initial conditions for each model compartment in the domain $[0, L]$, we adhere to the  following biologically relevant assumptions:
	\begin{enumerate}
		\item At the point of infection (POI) 
		$x=0$, the concentration of uninfected hepatocytes is very low, whereas the concentration of infected hepatocytes is extremely high.
		\item 
		At the POI, the initial concentrations of HBV capsids, viruses, SVPs and antibodies  are maximum and equal for all, \textit{i.e.}, $D_0=V_0=S_{v0}=A_0$.
	\end{enumerate}
	Following the similar  approach  as detailed in the article of Chaplain and Lolas \cite{chaplain2006mathematical},  the respective  initial conditions are formulated. In terms of mathematics, the initial concentrations are written in equation \eqref{eq:initial} and are visualized in Figure \ref{initial_condition_all}.
	\begin{equation} \label{eq:initial}
		\begin{aligned}
			X(x,0) &= M\left\{1 - \exp\left(-\frac{x^2}{\epsilon}\right)\right\}, \quad 0 \leq x \leq L, \\
			Y(x,0) &= M\exp\left(-\frac{x^2}{\epsilon}\right), \quad 0 \leq x \leq L, \\
			U(x,0) &= U_0\exp\left(-\frac{x^2}{\epsilon}\right), \quad 0 \leq x \leq L, \\
			V(x,0) &= V_0\exp\left(-\frac{x^2}{\epsilon}\right), \quad 0 \leq x \leq L, \\
			S_v(x,0) &= S_{v0}\exp\left(-\frac{x^2}{\epsilon}\right), \quad 0 \leq x \leq L, \\
			A(x,0) &= A_0\exp\left(-\frac{x^2}{\epsilon}\right), \quad 0 \leq x \leq L, \quad \text{with } \epsilon = 0.01.
		\end{aligned}
	\end{equation}
	Guided by the in vivo protocol in which
	infection takes place within the liver, it is  assumed that there is no-flux of capsids, viruses, SVPs and antibodies across the boundary of the liver, \textit{i.e.}, throughout the infection, the liver acts as an impregnable fortress, forbidding the entry   or exit of any particles. The end points of the liver are denoted by $x=0$ and $x=L$ in one-dimensional framework, shown in Figure \ref{liver}. These imposed  homogeneous Neumann  boundary conditions are represented by the following equations;
	\begin{equation}\label{eq:boundary_conditions_left}
		\frac{\partial U(x,t)}{\partial x}\Big|_{x=0}=\frac{\partial V(x,t)}{\partial x}\Big|_{x=0}=	\frac{\partial S_v(x,t)}{\partial x}\Big|_{x=0}=	\frac{\partial A(x,t)}{\partial x}\Big|_{x=0}=0,~t\geq 0,
	\end{equation}
	and 
	\begin{equation}\label{eq:boundary_conditions_right}
		\frac{\partial U(x,t)}{\partial x}\Big|_{x=L}=\frac{\partial V(x,t)}{\partial x}\Big|_{x=L}=	\frac{\partial S_v(x,t)}{\partial x}\Big|_{x=L}=	\frac{\partial A(x,t)}{\partial x}\Big|_{x=L}=0,~t\geq 0,
	\end{equation}
	where $\dfrac{\partial}{\partial x}$ denotes the outward normal derivative on the boundary of the domain  $x=0$ and $x=L$.\\
	\textbf{Note:} For simplicity, $X$, $Y$, $U$, $V$, $S_v$, and $A$ will be used instead of expressions $X(x,t)$, $Y(x,t)$, $U(x,t)$, $V(x,t)$, $S_v(x,t)$, and $A(x,t)$ throughout the rest of this study.

\section{Non-dimensional form of the proposed model}
\noindent
Due to the presence of high non-linearity in the proposed model problem,  the system of PDEs (equation \eqref{eq:Healthy cells}-\eqref{eq:antibody})  is solved numerically. 
In order to do non-dimensionalization  of the proposed  model, the following typical scales are used: 
\begin{enumerate}[(1)]
\item Reference length scale, $L$, is chosen to be the total length of the liver along the $x-$direction. The approximate value of $L$ is considered as $L=13.6~\mbox{cm}$ \cite{2019_Liver_size}.
\item Using $\mathcal{D}_l$, time scale can be written as $\bar{t}=\dfrac{L^2}{\mathcal{D}_l}$, where $\mathcal{D}_l$ is a reference chemical diffusion coefficient,
\item It is considered that  the total number of  liver cells  is denoted by $M=2\times 10^{11}$ \cite{2006_Murray}.
\item  Initially, maximum  values of concentration of capsids $D_0$, viruses $V_0$, SVPs  $S_0$, antibodies $A_0$,  are attained at the point of infection $x=0$. 
\end{enumerate}
By using these scales, the following non-dimensional variables and
parameters are obtained as follows:\\
\begin{itemize}
\item \textbf{Dimensionless model variables:} $t^{*}=\dfrac{t}{\bar{t}}$, $x^{*}=\dfrac{x}{L}$, $Y^{*}=\dfrac{Y}{M}$, $X^{*}=\dfrac{X}{M}$, $U^{*}=\dfrac{U}{U_0}$, $V^{*}=\dfrac{V}{V_0}$, $S^{*}=\dfrac{S_v}{S_{v0}}$ and $A^{*}=\dfrac{A}{A_0}$.
\item \textbf{Dimensionless model parameters:} $\lambda^{*}=\dfrac{\bar{t}\lambda}{M}$, $\mu^{*}=\mu\bar{t}$, $k^{*}=kV_0\bar{t}$, $\delta^{*}=\delta \bar{t}$, $a^{*}=\dfrac{aM\bar{t}}{U_0}$,  $\gamma^{*}=\gamma\bar{t}$, $\beta^{*}=\beta \bar{t}$,  $g^{*}=gA_0\bar{t}$, $\delta_v^{*}=\delta_v\bar{t}$,  $b^{*}=\dfrac{b\bar{t}M}{S_{v0}}$, $\delta_{s}^{*}=\delta_{s}\bar{t}$,  $\delta_a^{*}=\delta_{a}\bar{t}$, $h^{*}=h\bar{t}$, $r_a^*=r_A\bar{t}$, $A_m^*=\dfrac{A_m}{A_0}$, $\mathcal{D}_1^{*}=\dfrac{\mathcal{D}_1}{\mathcal{D}_l}$, $\mathcal{D}_2^{*}=\dfrac{\mathcal{D}_2}{\mathcal{D}_l}$, $\mathcal{D}_3^{*}=\dfrac{\mathcal{D}_3}{\mathcal{D}_l}$, $\mathcal{D}_4^{*}=\dfrac{\mathcal{D}_4}{\mathcal{D}_l}.$
\end{itemize}
At first, the equations \eqref{eq:Healthy cells}-\eqref{eq:antibody} are non-dimensionalized  using the aforementioned dimensionless variables and parameters. For the sake of  notational simplicity and clarity, omitting the star $(*)$ from each dimensionless variable and parameter, the non-dimensional governing equations are derived to be:
\begin{align}
\frac{\partial X}{\partial t}	&= \lambda-\mu X-kVX,
\label{eq:non_Healthy cells}\\
\frac{ \partial Y}{\partial t}	&=kVX-\delta Y ,
\label{eq:non_infected cells}\\
\frac{\partial U}{\partial t}&=\mathcal{D}_1\dfrac{\partial^2 U}{\partial x^2}+aY+\gamma(1-\alpha)U-\alpha\beta U-\delta U,
\label{eq:non_capsids}\\
\frac{\partial V}{ \partial t}&=\mathcal{D}_2\dfrac{\partial^2 V}{\partial x^2}+\alpha\beta U-g A V-\delta_v V ,
\label{eq:non_virus}\\
\frac{\partial S_v}{ \partial t}&=\mathcal{D}_3\dfrac{\partial^2 S_v}{\partial x^2} +bY-g A S_v-\delta_{s} S_v,
\label{eq:non_subviral}\\
\frac{\partial A}{\partial t}&=\mathcal{D}_4\dfrac{\partial^2 A}{\partial x^2}+h(V+S_v)+r_AA\left(1-\dfrac{A}{A_m}\right)-gA(S_v+V)-\delta_a A.
\label{eq:non_antigen}
\end{align}
Due to the non-dimensionalization, the domain of interest $[0,L]$ transforms to $[0,1]$.
The initial and boundary conditions take the following form in the  domain  $[0,1]$: \\
\begin{itemize}
\item \textbf{Non-dimensional form of initial conditions:}\\
\begin{equation} \label{eq:non_initial}
\left.
\begin{split}
&X(x,0)	= 1-\exp\left(-\frac{x^2}{\epsilon}\right),~ 0\leq x\leq 1,\\
&Y(x,0)	= \exp\left(-\frac{x^2}{\epsilon}\right),~ 0\leq x\leq 1,\\
&U(x,0)=\exp\left(-\frac{x^2}{\epsilon}\right),~ 0\leq x\leq 1,\\
&V(x,0)=\exp\left(-\frac{x^2}{\epsilon}\right),~ 0\leq x\leq 1,\\
&S_v(x,0)=\exp\left(-\frac{x^2}{\epsilon}\right),~ 0\leq x\leq 1,\\	
&A(x,0)=\exp\left(-\frac{x^2}{\epsilon}\right),~ 0\leq x\leq 1.		 	
\end{split}
\right\} 
\end{equation}
\item \textbf{Non-dimensional form of boundary conditions:}\\
\begin{equation}\label{eq:non_boundary_conditions}
\frac{\partial U}{\partial x}=\frac{\partial V}{\partial x}=	\frac{\partial S_v}{\partial x}=	\frac{\partial A}{\partial x}=0\;\;\text{on}\; \partial \Omega^* \times (0,+\infty),	
\end{equation}
where $\partial \Omega^*$	represents boundary  of $\Omega^*=[0,1]$ and $(0,+\infty)$ denotes the temporal domain.
\end{itemize}
\section{Numerical method} \label{Section: Numerical method}
\noindent
The computational  domain $[0,1]$ is divided into $(N-1)$ equal sub-intervals $0=x_0\le x_1\le x_2\le \cdots\le x_{i-1}\le x_i\le x_{i+1}\le\cdots\le x_{N-2}\le x_{N-1}=1$	 with step length $\Delta x=\dfrac{1}{(N-1)}$. Let $\Delta t$ be the time step length and $n$ be the step number to reach the final time $T_f$ i.e., $T_f=n\Delta t$. Here, the central difference scheme for spatial derivatives and the forward scheme for temporal derivatives (FTCS) is used to discretize the governing equations \eqref{eq:non_Healthy cells}-\eqref{eq:non_antigen}. Using the following notations: 
\begin{align*}
	X(x,t)&=X(i\Delta x, n\Delta t)=X_i^n,\\
	Y(x,t)&=Y(i\Delta x, n\Delta t)=Y_i^n,\\
	U(x,t)&=U(i\Delta x, n\Delta t)=U_i^n,\\
	V(x,t)&=V(i\Delta x, n\Delta t)=V_i^n,\\
	S_v(x,t)&=S_v(i\Delta x, n\Delta t)={S_v}_{i}^n,\\
	A(x,t)&=A(i\Delta x, n\Delta t)=A_i^n,
\end{align*}
 the corresponding discretized forms of the equations \eqref{eq:non_Healthy cells}-\eqref{eq:non_antigen} at the interior nodes  can be written  as
\begin{align}
X_i^{n+1}&=X_i^n+\lambda\Delta t-\mu\Delta t X_i^{n}-k\Delta t V_i^nX_i^n,\\
Y_i^{n+1}&=Y_i^n-\delta\Delta t Y_i^{n}+k\Delta t V_i^nX_i^n,\\
-p_1U_{i-1}^{n+1}+(1+2p_1)U_{i}^{n+1}-p_1U_{i+1}^{n+1}&=U_{i}^{n}+a\Delta t Y_i^n+\gamma \Delta t (1-\alpha)U_i^n-\alpha\beta\Delta t U_i^n-\delta\Delta U_i^n,
\label{eq:discre_capsid} \\
-p_2V_{i-1}^{n+1}+(1+2p_2)V_{i}^{n+1}-p_2V_{i+1}^{n+1}&=V_{i}^{n}+\alpha\beta\Delta t U_i^n-g\Delta t A_i^n V_i^n-\delta_v\Delta t V_i^n,
\label{eq:discre_virus}\\
-p_3{S_v}_{i-1}^{n+1}+(1+2p_3){S_v}_{i}^{n+1}-p_3{S_v}_{i+1}^{n+1}&={S_v}_{i}^{n}+b\Delta tY_i^n-g\Delta t {S_v}_{i}^nA_i^n-\delta_{s}\Delta t {S_v}_{i}^n,
\label{eq:discre_sub_viral}\\
-p_4A_{i-1}^{n+1}+(1+2p_4)A_{i}^{n+1}-p_4A_{i+1}^{n+1}&=A_{i}^{n}+(h-gA_i^n)({S_v}_{i}^n+V_i^n)+r_AA_i^n\left(1-\frac{A_i^n}{A_m}\right)-\delta_{a}\Delta t A_{i}^n,
\label{eq:discre_antibody}
\end{align}
where $i=1,2,\cdots, N-2, n=0,1,2,\cdots.$ Here, $ p_1=\dfrac{\Delta t }{\Delta x^2}\mathcal{D}_1$, $p_2=\dfrac{\Delta t }{\Delta x^2}\mathcal{D}_2$, $p_3=\dfrac{\Delta t }{\Delta x^2}\mathcal{D}_3$, $p_4=\dfrac{\Delta t }{\Delta x^2}\mathcal{D}_4$.  
\noindent	
Using ghost point criteria at the boundary nodes, \textit{i.e.}, $i=0,~\mbox{and}~i=N-1$, equations \eqref{eq:discre_capsid},\eqref{eq:discre_virus}, \eqref{eq:discre_sub_viral} and \eqref{eq:discre_antibody} take  the following forms:

\noindent
\begin{itemize}
\item  \textbf{At left boundary ($x=0$):}
\begin{align}
(1+2p_1)U_0^{n+1}-2p_1U_1^{n+1}&=U_0^n+a\Delta t Y_0^n+\gamma \Delta t (1-\alpha)U_0^n-\alpha\beta\Delta t U_0^n-\delta\Delta U_0^n,\\
(1+2p_2)V_{0}^{n+1}-2p_2V_{1}^{n+1}&=V_{0}^{n}+\alpha\beta\Delta t U_0^n-g\Delta t A_0^n V_0^n-\delta_v\Delta t V_0^n,\\
(1+2p_3){S_v}_{0}^{n+1}-2p_3{S_v}_{1}^{n+1}&={S_v}_{0}^{n}+b\Delta tY_0^n-g\Delta t {S_v}_{0}^nA_0^n-\delta_{s}\Delta t {S_v}_{0}^n,\\
(1+2p_4)A_{0}^{n+1}-2p_4A_{1}^{n+1}&=A_{0}^{n}+(h-gA_0^n)({S_v}_{0}^n+V_0^n)+r_AA_0^n\left(1-\frac{A_0^n}{A_m}\right)-\delta_{a}\Delta t A_{0}^n.
\end{align}
\item \textbf{At right boundary ($x=1$):}
\begin{align}
-2p_1U_{N-2}^{n+1}+(1+2p_1)U_{N-1}^{n+1}&=U_{N-1}^{n}+a\Delta t I_{N-1}^n+\gamma \Delta t (1-\alpha)U_{N-1}^n-\alpha\beta\Delta t U_{N-1}^n-\delta\Delta U_{N-1}^n,\\
-2p_2V_{N-2}^{n+1}+(1+2p_2)V_{N-1}^{n+1}&=V_{N-1}^{n}+\alpha\beta\Delta t U_{N-1}^n-g\Delta t A_{N-1}^n V_{N-1}^n-\delta_v\Delta t V_{N-1}^n,\\
-2p_3{S_v}_{N-2}^{n+1}+(1+2p_3){S_v}_{N-1}^{n+1}&={S_v}_{N-1}^{n}+b\Delta tY_{N-1}^n-g\Delta t {S_v}_{N-1}^nA_i^n-\delta_{s}\Delta t {S_v}_{N-1}^n,\\
-2p_4A_{N-2}^{n+1}+(1+2p_4)A_{N-1}^{n+1}&=A_{N-1}^{n}+(h-gA_{N-1}^n)({S_v}_{N-1}^n+V_{N-1}^n)\nonumber\\&+r_AA_{N-1}^n\left(1-\frac{A_{N-1}^n}{A_m}\right)-\delta_{a}\Delta t A_{N-1}^n.
\end{align}
\end{itemize}
The corresponding systems of equations for $U$, $V$, $S_v$ and $A$ are solved using the well-known tridiagonal matrix algorithm (TDMA).
\section{Parameter estimation}
\noindent	
The baseline values of some parameters  are directly taken  from existing  literature. It is assumed that the infectious viral particles and SVPs  degrade naturally with same rate, \textit{i.e.}, $\delta_v=\delta_{s}$. The value of volume fraction of  rcDNA-containing capsids ($\alpha$) in favor of virus production are considered as 0.8 throughout this study. Since, the ratio of Dane particles to SVPs ranges from 1:10,000 to 1:100,000 \cite{2013_gerlich_medical,2015_luckenbaugh_genome,2017_hu_complete}, it is noted that the production rate of SVPs is accounted to be at least 1000 times that of virus production rate, \textit{i.e.}, $b=1000a$.
The diffusion coefficients ($\mathcal{D}_1, \mathcal{D}_2,\mathcal{D}_3,\mathcal{D}_4$) are systematically varied within a realistic range to capture the diverse  spectrum of viral infection dynamics. All details regarding the parameters, such as descriptions, units, values, and their corresponding references are mentioned in Table \ref{Table-2}.

\begin{table}[h!]
\scriptsize
\caption{ \label{Table-2} Descriptions of model variables and parameters.}
\begin{tabular}{|c|l|l|l|l|c|}
\hline
\cellcolor{blue!30}Variables & \multicolumn{5}{l|}{\cellcolor{blue!30}Descriptions} \\ \hline
$X$ & \multicolumn{5}{l|}{Number of uninfected hepatocytes} \\ \hline
$Y$ & \multicolumn{5}{l|}{Number of infected hepatocytes} \\ \hline
$U$ & \multicolumn{5}{l|}{Number of HBV DNA-containing capsids} \\ \hline
$V$ & \multicolumn{5}{l|}{Number of viruses} \\ \hline
$S_v$ & \multicolumn{5}{l|}{Number of SVPs} \\ \hline
$A$ & \multicolumn{5}{l|}{Number of  antibodies} \\ \hline
\cellcolor{blue!30}Parameters & \multicolumn{1}{l|}{\cellcolor{blue!30}Descriptions} & \multicolumn{1}{c|}{\cellcolor{blue!30}Dimensional values} & \multicolumn{1}{c|}{\cellcolor{blue!30}Units} & \multicolumn{1}{c|}{\begin{tabular}[c]{@{}c@{}}\cellcolor{blue!30}Non-dimensional\\ \cellcolor{blue!30} values\end{tabular}} &\cellcolor{blue!30} Source \\ \hline
$\lambda$  & \multicolumn{1}{l|}{Birth rate of uninfected hepatocytes } & \multicolumn{1}{c|}{$2.6\times10^7$} & \multicolumn{1}{c|}{$\mbox{cells  ml}^{-1}\mbox{day}^{-1}$} & \multicolumn{1}{c|}{0.0084} & \cite{2007_Dahari} \\ \hline
$\mu$		& \multicolumn{1}{l|}{Death rate of uninfected hepatocytes}				& \multicolumn{1}{c|}{$0.01$}     & \multicolumn{1}{l|}{$\mbox{day}^{-1}$}   &\multicolumn{1}{c|}{0.8438} & \cite{2007_Dahari}	   			\\ \hline
$k$  & \multicolumn{1}{l|}{Virus-to-cell infection rate		} & \multicolumn{1}{c|}{$1.67\times10^{-12}$} & \multicolumn{1}{c|}{$\mbox{ml virion}^{-1}\mbox{day}^{-1}$} & \multicolumn{1}{c|}{1.4091} & \cite{2006_Murray} \\ \hline
$a$  & \multicolumn{1}{l|}{Production rate of capsid from} & \multicolumn{1}{c|}{150} & \multicolumn{1}{c|}{$\mbox{capsids cell}^{-1}\mbox{day}^{-1}$} & \multicolumn{1}{c|}{253125} & \cite{2006_Murray} \\ 
& \multicolumn{1}{l|}{infected hepatocytes} & \multicolumn{1}{c|}{} & \multicolumn{1}{c|}{} & \multicolumn{1}{c|}{} & \\
\hline	
$b$			& \multicolumn{1}{l|}{Production rate of SVPs from} 	& \multicolumn{1}{c|}{$1000a$}      & \multicolumn{1}{c|}{$\mbox{capsids cell}^{-1}\mbox{day}^{-1}$}   & \multicolumn{1}{c|}{1468} &\cite{2006_Murray}		\\
&  \multicolumn{1}{l|}{infected hepatocytes}	& \multicolumn{1}{c|}{}     & \multicolumn{1}{c|}{}    & \multicolumn{1}{c|}{}&	\\
\hline
$\beta$ 	&  \multicolumn{1}{l|}{Virus production rate from capsids}				&  \multicolumn{1}{c|}{0.87}    &   \multicolumn{1}{c|}{$\mbox{day}^{-1}$} &  \multicolumn{1}{c|}{73.40} &\cite{2006_Murray}				\\
\hline	
$\delta$	& \multicolumn{1}{l|}{Death rate of infected hepatocyte}  & \multicolumn{1}{c|}{0.053}     &  \multicolumn{1}{c|}{$\mbox{day}^{-1}$}  & \multicolumn{1}{c|}{4.47} &\cite{2006_Murray}\\
&  \multicolumn{1}{l|}{as well as capsids} &\multicolumn{1}{c|}{ }     &\multicolumn{1}{c|}{}   &\multicolumn{1}{c|}{}   &					\\ 
\hline
$\delta_v$ 		& \multicolumn{1}{l|}{Death rate of viruses}						& \multicolumn{1}{c|}{3.8}     &  \multicolumn{1}{c|}{$\mbox{day}^{-1}$}  & \multicolumn{1}{c|}{320} &\cite{2006_Murray}						\\
\hline		
$\eta$ 	& \multicolumn{1}{l|}{Volume fraction of HBV}  			& \multicolumn{1}{c|}{$0\leq \alpha\leq 1$}      & \multicolumn{1}{c|}{Unitless}  &	\multicolumn{1}{c|}{0.8}	&	\\
&   \multicolumn{1}{l|}{DNA-containing capsids} 			&\multicolumn{1}{c|}{}  &\multicolumn{1}{c|}{}    &\multicolumn{1}{c|}{}    &			\\
\hline
$\gamma$ 	& \multicolumn{1}{l|}{Recycling rate	}		&  \multicolumn{1}{c|}{0.6931}     &\multicolumn{1}{c|}{$\mbox{day}^{-1}$}  & \multicolumn{1}{c|}{58.21}& \cite{2015_Murray}			\\
\hline
$g$ 	& \multicolumn{1}{l|}{Binding rate of antibody and virus }	&   \multicolumn{1}{c|}{$10^{-4}$  } &\multicolumn{1}{c|}{$\mbox{day}^{-1}$ }  & \multicolumn{1}{c|}{8437} &\cite{2017_meskaf_optimal}\\
& \multicolumn{1}{l|}{ or	SVPs	}	&   \multicolumn{1}{c|}{} &\multicolumn{1}{c|}{ }  & \multicolumn{1}{c|}{} &\\
\hline
$h$ 	& \multicolumn{1}{l|}{production rate of antibodies}			&  \multicolumn{1}{c|}{0.01}    & \multicolumn{1}{c|}{$\mbox{day}^{-1}$}   & \multicolumn{1}{c|}{8437500}& \cite{2017_meskaf_optimal}\\
\hline
$\delta_{s}$ 	& \multicolumn{1}{l|}{Death rate of sub-viral particles}			&  \multicolumn{1}{c|}{3.8 }   & \multicolumn{1}{c|}{$\mbox{day}^{-1}$}  & \multicolumn{1}{c|}{320} & Assumed			\\
\hline
$\delta_{a}$ 	& \multicolumn{1}{l|}{Death rate of antibodies}			&    \multicolumn{1}{c|}{0.15}  & \multicolumn{1}{c|}{$\mbox{day}^{-1}$}  & \multicolumn{1}{c|}{16.8} & \cite{2017_meskaf_optimal}			\\
\hline
$\mathcal{D}_1$ 	& \multicolumn{1}{l|}{Diffusion coefficient of capsids	}		& \multicolumn{1}{c|}{0.08}      & \multicolumn{1}{c|}{$cm^2\mbox{day}^{-1}$} & \multicolumn{1}{c|}{0.0370} & -			\\
\hline
$\mathcal{D}_2$ 	&  \multicolumn{1}{l|}{Diffusion coefficient of viruses	}			&  \multicolumn{1}{c|}{0.09}     & \multicolumn{1}{c|}{$cm^2\mbox{day}^{-1}$}  & \multicolumn{1}{c|}{0.0417} & -			\\
\hline
$\mathcal{D}_3$ 	&  \multicolumn{1}{l|}{Diffusion coefficient of SVPs}				&  \multicolumn{1}{c|}{0.09}     & \multicolumn{1}{c|}{$cm^2 \mbox{day}^{-1}$}  & \multicolumn{1}{c|}{0.0417} & -			\\
\hline
$\mathcal{D}_4$ 	&  \multicolumn{1}{l|}{Diffusion coefficient of antibodies}				&   \multicolumn{1}{c|}{0.08 }   & \multicolumn{1}{c|}{$cm^2\mbox{day}^{-1}$ }  & \multicolumn{1}{c|}{0.0370} & 	-		\\				
\hline
\end{tabular}
\end{table}
\section{Numerical experiment}
\noindent
By applying the methodology described  in Section \ref{Section: Numerical method}, the  dimensionless system of equations \eqref{eq:non_Healthy cells}-\eqref{eq:non_antigen} is solved considering the above-mentioned initial conditions \eqref{eq:non_initial} and boundary conditions \eqref{eq:non_boundary_conditions}. 
However, for the sake of illustration, the obtained results are presented in dimensional form to provide a clear understanding about the actual biology and to facilitate comparison between different scenarios.
\subsection{Effects of diffusion}
\noindent
In general, the mobility of infectious virus particles inside the host plays some decisive roles in the progression and transmission of any kind of viral infection. In this section, in order to examine the relative importance of diffusion of capsids, viruses, SVPs and antibodies on HBV infection dynamics,  we first consider the system in the absence of any kind of diffusion,( \textit{i.e.,} $\mathcal{D}_1=\mathcal{D}_2=\mathcal{D}_3=\mathcal{D}_4=0$). Subsequently, when diffusivity is imposed into the model,   the values of diffusion coefficients ($\mathcal{D}_1,\mathcal{D}_2,\mathcal{D}_3,\mathcal{D}_4$) are varied within the empirically reliable range to explore diverse scenarios and uncover testable biological aspects. In this case, none of the diffusion coefficients are set to zero. We proceed in the following manner:

\begin{enumerate}
\item The effects of diffusion on uninfected and infected hepatocytes.
\item The effects of diffusion on HBV capsids and free viruses.
\item The effects of diffusion on SVPs and antibodies.
\end{enumerate}

\subsubsection{Effects of diffusion on uninfected and infected hepatocytes}
\begin{figure}[h!]
\centering
\includegraphics[width=17cm, height=11cm]{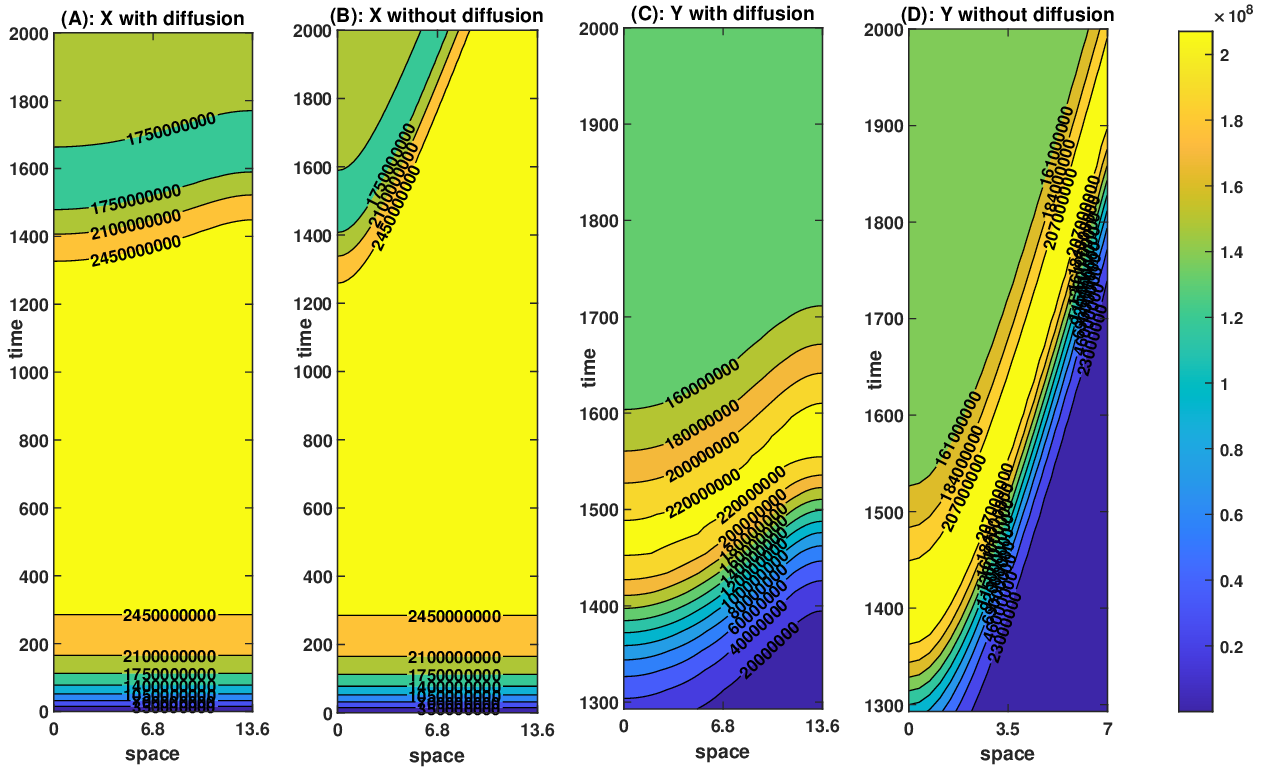}
\caption{Contour color plots: The effects of diffusion on uninfected and infected hepatocytes.}
\label{fig: Uninfected and Infected contour}
\end{figure}

The changes in the dynamics of uninfected and infected hepatocytes are shown in the Figure \ref{fig: Uninfected and Infected contour}. In order to visualize the concentration profiles of respective classes with respect to time and space in a easier way,  contour color plots are presented.  Figure \ref{fig: Uninfected and Infected contour}A and Figure \ref{fig: Uninfected and Infected contour}B display the distribution pattern of uninfected hepatocytes with and without considering the diffusivity of HBV capsids, viruses, SVPs and antibodies, respectively. Significant variations are observed in the distribution patterns. At the early stages of infection, the influence of diffusion on the uninfected hepatocytes compartment is not  noticeable, as the concentration remains nearly uniform across times (contour lines are parallel to space-axis). However, subsequent to this initial phase, diffusion begins to exert substantial impacts. For these set of parameters, after the time $t\approx 1330^{th}$ day, the distribution of uninfected hepatocytes shows a paradigm shift in the contour line (for example, contour line 2450000000). In the diffusion model, uninfected hepatocytes exhibit an almost uniform distribution along space, whereas in the non-diffusion model, they display a non-uniform pattern. In the non-diffusion model, the infection takes a considerable amount of time to spread throughout the liver.

In case of  infected hepatocytes, similar kind of dynamics are observed  in Figure \ref{fig: Uninfected and Infected contour}C and Figure \ref{fig: Uninfected and Infected contour}D as shown in uninfected hepatocytes. In the non-diffusion model, the rate of temporal changes in the concentration of infected hepatocytes is higher compared to diffusion one. 
In essence, the diffusion of virion components (capsids, virions, SVPs) and antibodies enhances the infection increasing the susceptibility in cells. 
However, both models lead to the same steady state for both type of hepatocytes.

%

\subsubsection{Effects of diffusion on HBV capsids and free viruses}
\begin{figure}[h]
\centering
\includegraphics[width=17cm, height=11cm]{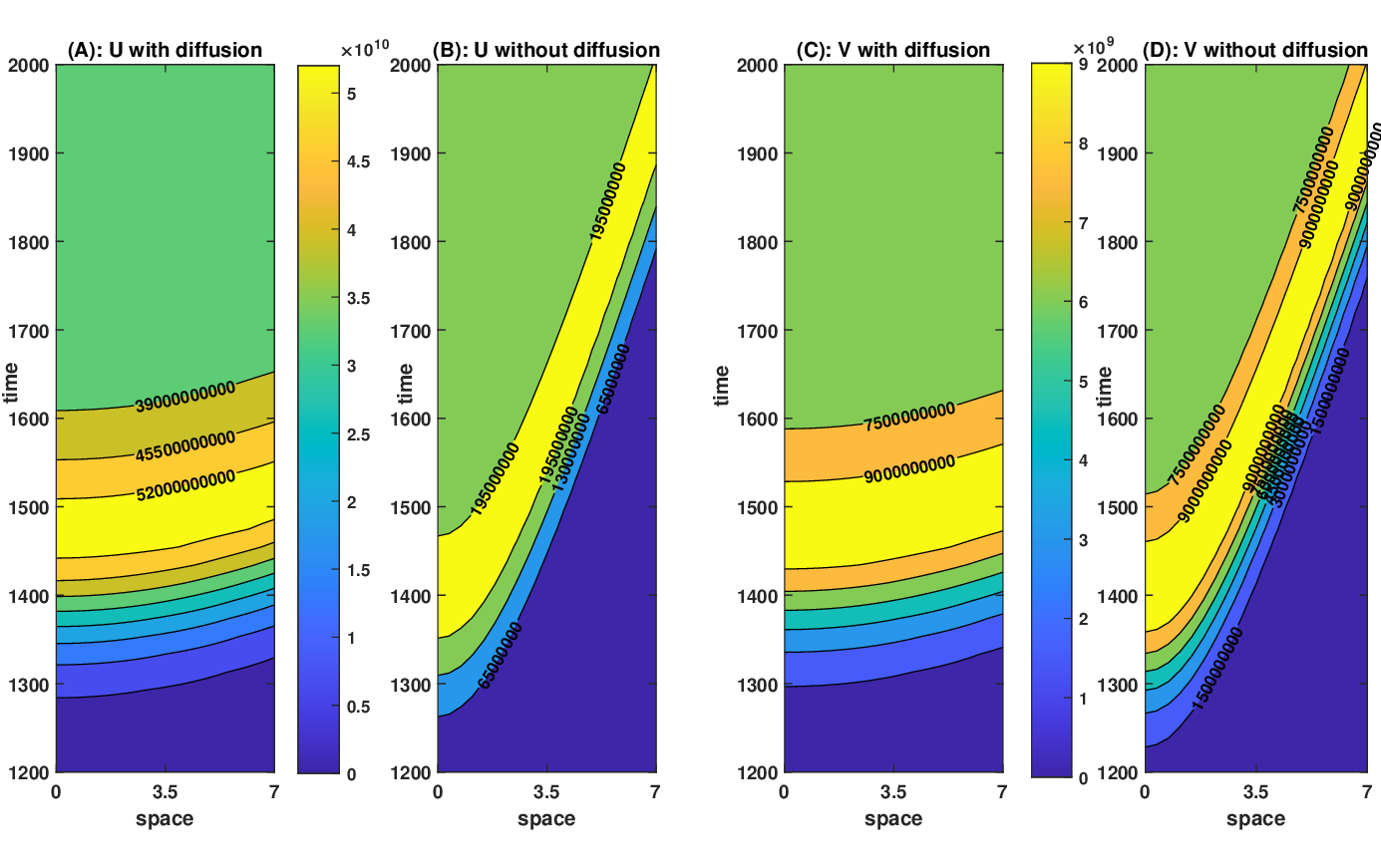}
\caption{Contour color plots: The effects of diffusion on HBV capsids and infectious virus particles.}
\label{fig: Capsids and Viruses contour}
\end{figure}

The contour plots shown in Figure \ref{fig: Capsids and Viruses contour}  delineates the concentration profiles of HBV DNA-containing capsids and viruses. For non-zero values of diffusion coefficients $\mathcal{D}_1, \mathcal{D}_2, \mathcal{D}_3, \mathcal{D}_4$, the densities of capsids and viruses demonstrate an nearly orthotropic pattern (property changes in mutually perpendicular directions) in Figure \ref{fig: Capsids and Viruses contour}A (capsids) and Figure \ref{fig: Capsids and Viruses contour}C (viruses). It is also observed that for same time (suppose $t=1400^{th}$day), there are significant differences in the concentrations  of capsids and viruses compartments between the diffusion and non-diffusion models.  In the absence of diffusion, the spread of capsids and virion particles occurs slowly (Figure \ref{fig: Capsids and Viruses contour}B and Figure \ref{fig: Capsids and Viruses contour}D). However, due to the inclusion of diffusions, viruses start to propagate rapidly throughout the liver, leading to a spatially homogeneous distribution. Therefore, diffusivities of capsids, viruses, SVPs and antibodies influence the infection pattern and its dynamics.

\subsubsection{Effects of diffusion on SVPs and antibodies}
\begin{figure}[h]
\centering
\includegraphics[width=17cm, height=11cm]{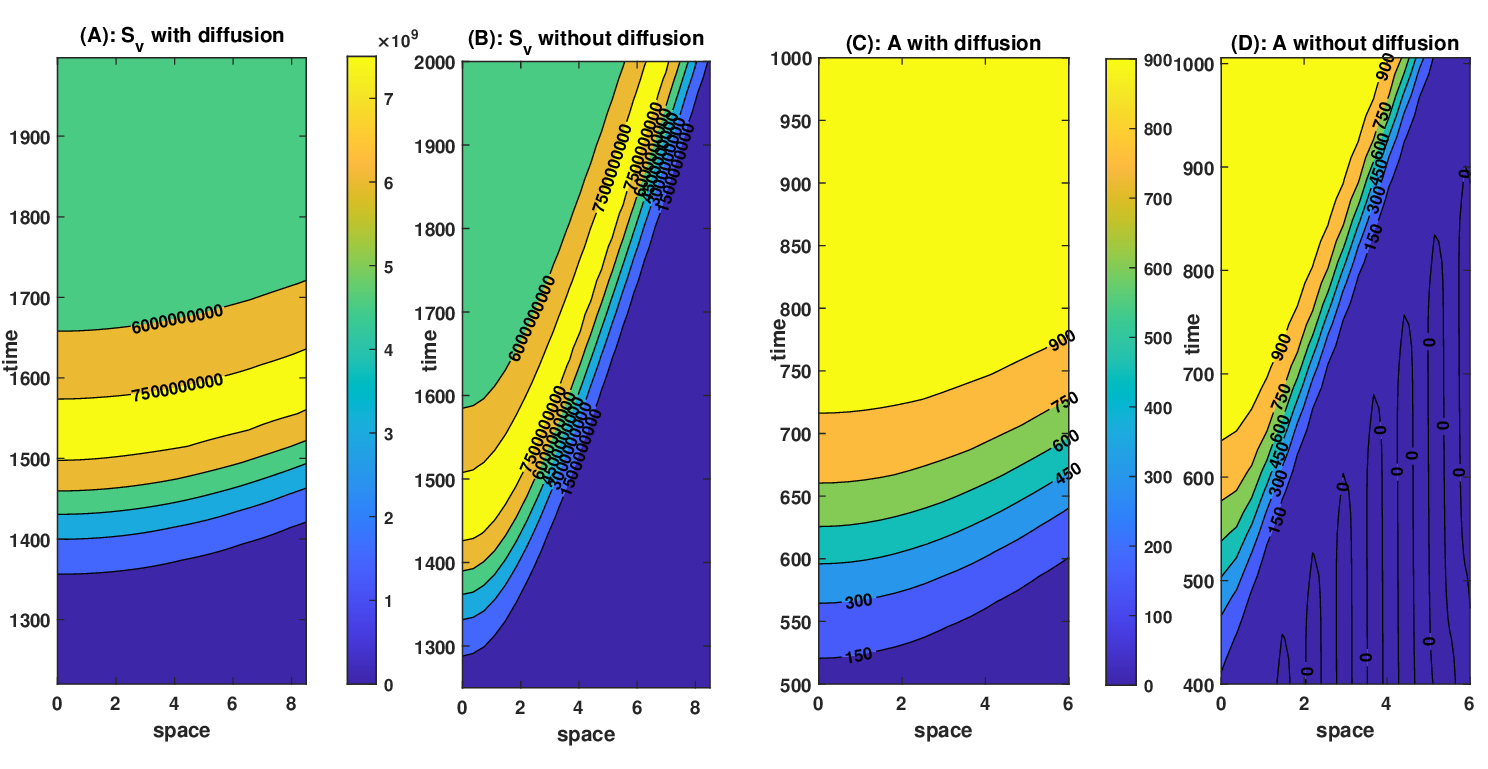}
\caption{Contour color plots: The effects of diffusion on SVPs and antibodies.}
\label{fig: SVPs and Antibody contour}
\end{figure}

In this section, we discuss how the spatial movement of capsids, viruses, SVPs, and antibodies affects the concentrations of non-infectious SVPs and virus-neutralizing antibodies. In Figure \ref{fig: SVPs and Antibody contour}A and Figure \ref{fig: SVPs and Antibody contour}C, the values of SVPs $(S_v(x,t))$ and antibodies $(A(x,t))$ with diffusion are visualized using contour color plots. On the other hand, Figure \ref{fig: SVPs and Antibody contour}B and Figure \ref{fig: SVPs and Antibody contour}D shows the corresponding concentrations of SVPs and antibodies under non-diffusion conditions.  Initially, it is considered that the infection starts from the left side of the liver, \textit{i.e.}, from the origin $(0,0)$ according to our $xt-$frame. There are notable changes in the concentrations of SVPs and antibodies observed exclusively on the left side of the liver, while the right side exhibits an absence of both SVPs and antibodies in the non-diffusion model. The  antibodies are concentrated only near the site of infection in non-diffusion model. It is quite impractical.  While diffusion is taken into account, substantial variations are observed occurring in both sides of the liver.   \\

\noindent
Contour plots shown in Figure \ref{fig: Uninfected and Infected contour}, \ref{fig: Capsids and Viruses contour} and \ref{fig: SVPs and Antibody contour} indicate that  all  compartments (uninfected hepatocytes, infected hepatocytes, capsids, viruses, SVPs and antibodies) reach an endemic steady-state more rapidly in case of non-diffusion model compared to the diffusion one.
\subsection{Effects of SVPs }
\begin{figure}[h!]
\centering
\includegraphics[width=15cm, height=14cm]{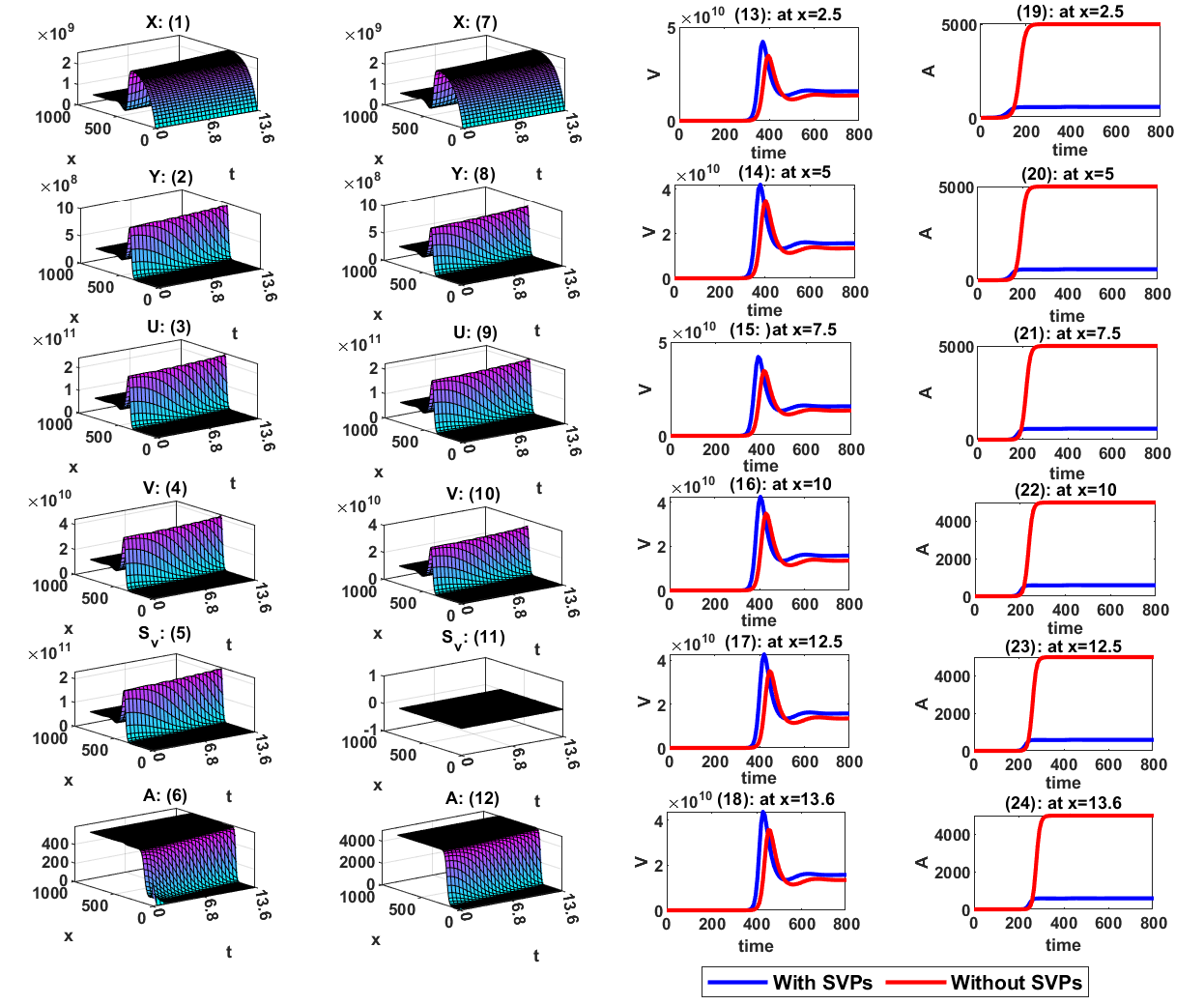}
\caption{Comparisons between the virus and antibody profiles considering both the cases:  with and without SVPs. Here, the blue lines illustrate the system's solution incorporating SVPs, while the red curves depict the solution without SVPs.}
\label{fig:virus_and_antibody_with_and_without_svps_24}
\end{figure}

\noindent
SVPs are one of the intracellular viral components that weaken the host immune system. These viral elements release from the infected hepatocytes in a large number (typically 1,000- to 100,000-fold) compare to infectious virus particles during infection period \cite{2013_gerlich_medical,2015_luckenbaugh_genome,2017_hu_complete}. In this section, we investigate the effects of  SVPs on virus and antibody growth dynamics throughout the liver. The system of equations \eqref{eq:non_Healthy cells} \eqref{eq:non_antigen} is solved  by considering both the cases: (i) with SVPs and (ii) without SVPs.  In Figure \ref{fig:virus_and_antibody_with_and_without_svps_24} (1-6), the solution for  the first case \textit{i.e.} solutions with the SVPs are displayed. On the other hand, the outcomes for the second case (without SVPs) are depicted in Figure \ref{fig:virus_and_antibody_with_and_without_svps_24} (7-12). 
Upon comparing the solutions for both cases, notable changes are observed, particularly in the virus and antibody profiles.
The snapshots of the virus  and antibody compartments are captured at different  positions, such as at $x=2.5,~x=5,~x=7.5,~x=10,~x=12.5$ and $x=13.6$, and  shown in Figure \ref{fig:virus_and_antibody_with_and_without_svps_24} (13-18) for virus and Figure \ref{fig:virus_and_antibody_with_and_without_svps_24} (19-24) for antibody. Due to the integration of SVPs into the model, these following critical insights are noted:
\begin{enumerate}
\item  There is a discernible rise in the peak of virus compartment (refers Figure \ref{fig:virus_and_antibody_with_and_without_svps_24} (13-18)).
\item The virus compartment stabilizes at a higher level  (refers Figure \ref{fig:virus_and_antibody_with_and_without_svps_24} (13-18)).
\item The peak level for the virions compartment rises quickly (refers Figure \ref{fig:virus_and_antibody_with_and_without_svps_24} (13-18)).
\item  After a certain time period, substantial changes are observed in the solution for the antibody compartment (refers Figure \ref{fig:virus_and_antibody_with_and_without_svps_24} (19-24)). The stability level of antibodies significantly decreases due to the presence of SVPs.
The primary reason is the excessive production of SVPs which leads to the depletion of a considerable amount of antibodies, resulting in a swift decline in antibody concentration.
\end{enumerate}
Therefore, the above findings  suggest that SVPs play some important roles in development of this viral infection. Similar types of results have been reported in the literature. For example, Bruns et al. \cite{1998_bruns_enhancement} found that SVPs significantly enhance intracellular viral replication and gene expression. Recently, Rydell et al. \cite{2017_rydell_hepatitis} confirmed that  SVPs
significantly reduced the neutralization of virus particles by antibodies. The removal of spherical SVPs stands as an important milestone in achieving the functional cure of HBV infection \cite{2020_vaillant_hbsag}. Thus, in conclusion, SVPs merit attention as a promising avenue for future therapeutic strategies targeting viral agents.
%
\subsection{Effects of capsid recycling on the production of SVPs}
\begin{figure}[h!]
\centering
\includegraphics[width=18cm, height=18cm]{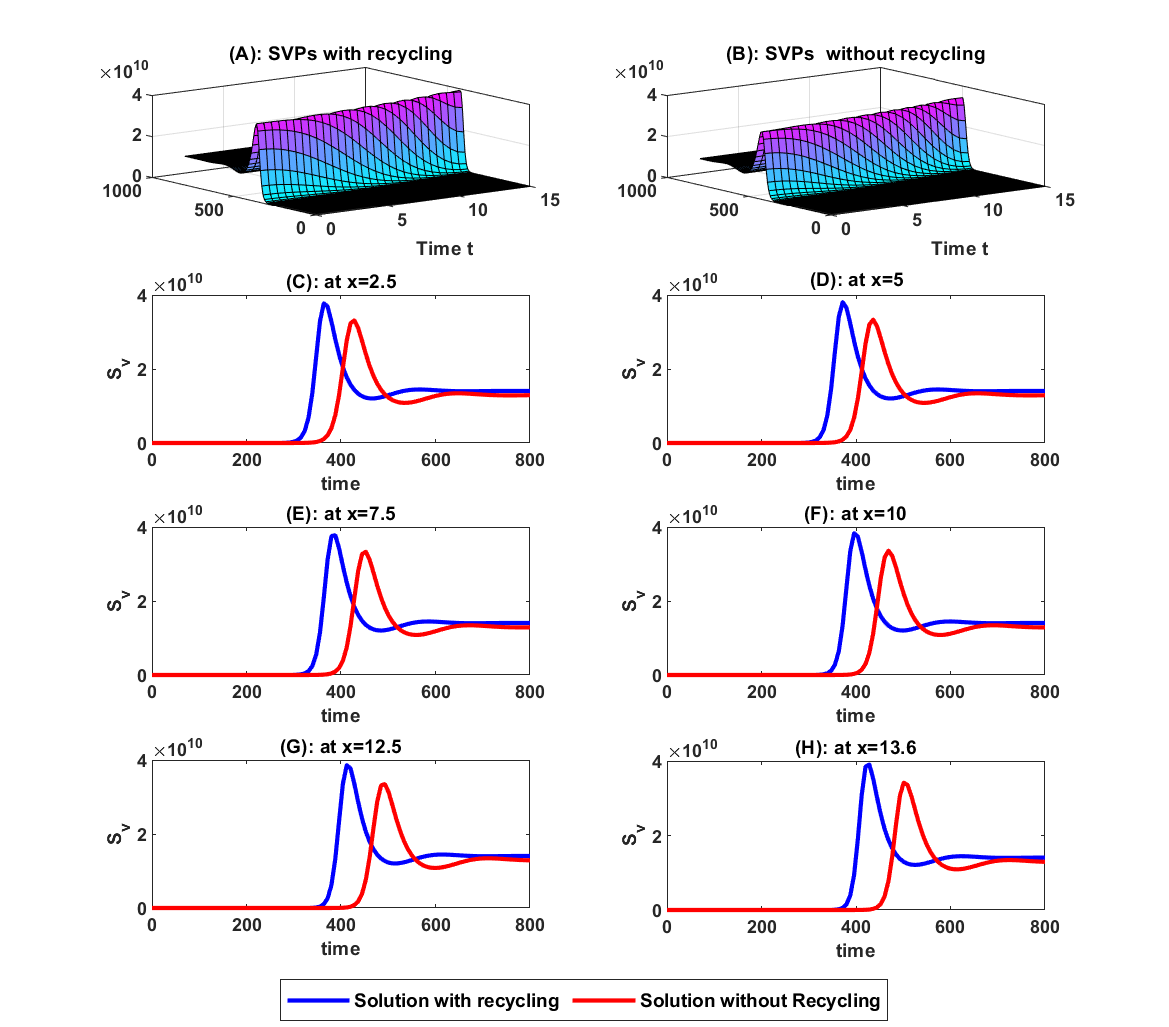}
\caption{The effects of recycling on SVPs. Figure \ref{The effect of recycling on SVPs} (A): surface plot of SVPs when recycling is taken into account. Figure \ref{The effect of recycling on SVPs} (B): surface plot of SVPs when recycling is not considered. Figure \ref{The effect of recycling on SVPs} (C)-Figure \ref{The effect of recycling on SVPs} (H): The comparison between the concentration of SVPs at $x=2.5$, $x=5$, $x=7.5$, $x=10$, $x=12.5$, and $x=15$, with respect to time. Here, blue curves represent the solution of the system when recycling of capsids is considered. On the other flip, red curves denotes the solutions when recycling effects are ignored.}
\label{The effect of recycling on SVPs}
\end{figure}

\noindent
The recycling of capsids acts as a crucial mechanism in  HBV. Recently, Sutradhar and Dalal \cite{2023_sutradhar_recycling,2023_sutradhar_fractional} found how the recycling pathway substantially augments viral concentration during HBV infection.  To the best of our knowledge, we didn't find any prior documented investigations  related the  interplay between capsid recycling and SVPs in the literature. The possible associations  between capsid recycling and SVPs are  studied here. The proposed model is solved under two distinct cases:\vspace{0.2cm}\\
\textbf{ Case-1: without recycling ($\gamma=0,~\alpha=1$)}\\
In this case, the impacts of capsid recycling is ignored, \textit{i.e.}, the simulation is done by considering $\gamma=0$ and the corresponding profiles of SVPs are shown in Figure \ref{The effect of recycling on SVPs}A.\vspace{0.2cm}\\
\textbf{ Case-2: with recycling ($\gamma\neq0,~\alpha<1$)} \\
We incorporate the recycling effects into the proposed model through the term $\gamma(1-\alpha)U(x,t)$ in equation \eqref{eq:capsids}. By assigning a realistic estimated value to $\gamma$, we solve the system of equations. The resulting surface plot for SVPs is depicted in Figure \ref{The effect of recycling on SVPs}B.\\ 
In Figure \ref{The effect of recycling on SVPs}(C)-Figure \ref{The effect of recycling on SVPs}(H), the solutions of SVPs for various positions are highlighted. The trends of the solutions exhibit similarities but demonstrate significant differences in values. The extreme values for each solution vary considerably. Incorporation of recycling of capsids results in a higher peak in the solution compared to when recycling is not considered. Therefore, the recycling of capsids is implicated in enhancing the concentration of SVPs during HBV infection.
\section{Infection initiating from  multiple points on the liver} 
\begin{figure}[h!]
\centering
\includegraphics[width=12cm, height=5cm]{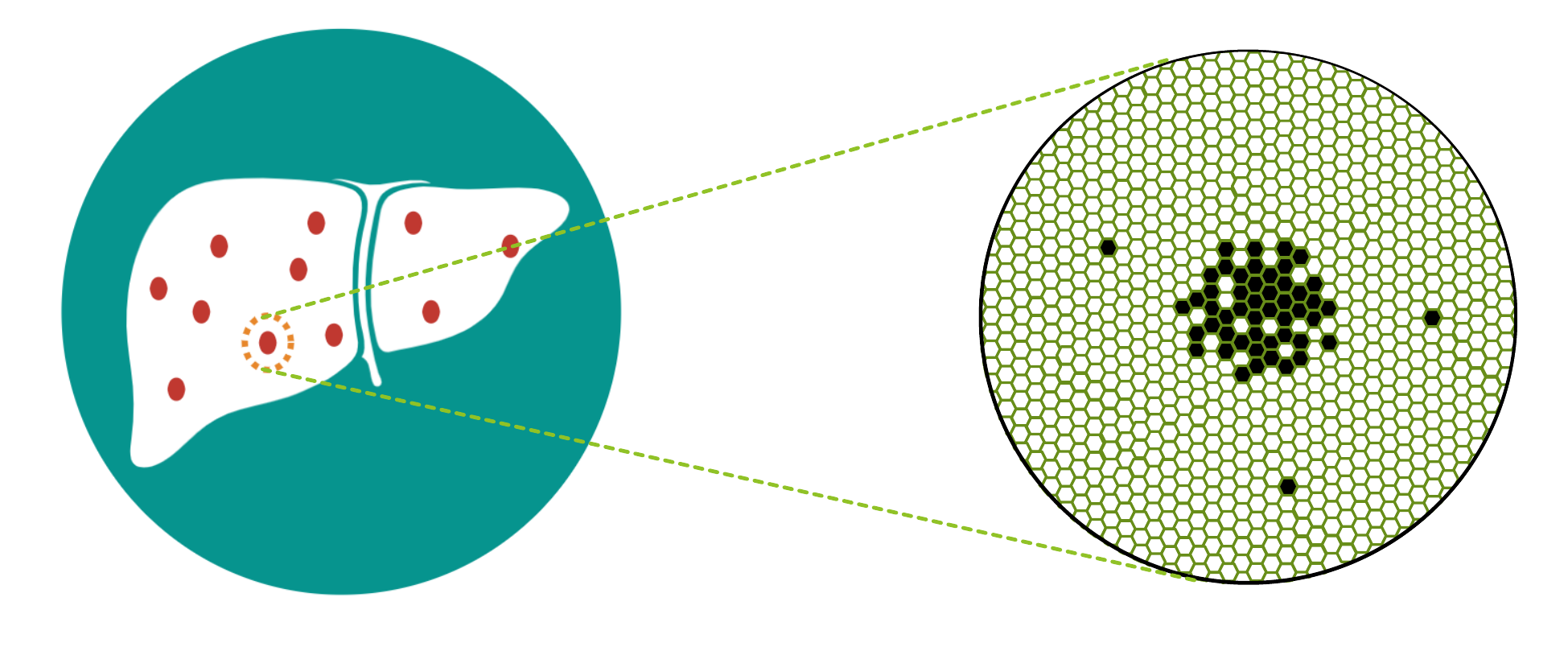}
\caption{Distribution of infection in the liver during the course of HBV infection.}
\label{fig: Distribution of infection}
\end{figure}

\noindent
In the preceding sections, we made the assumption that  the infection initially originates only from a single point at $x=0$ (left side). Subsequently,  the corresponding initial conditions were considered as described in the equation \eqref{eq:initial}. This virus predominantly infects hepatocytes which serve as the primary functional cells within the liver. However, the infection doesn't necessarily involve every hepatocyte in the liver at same time. The virus may infect hepatocytes in various parts of the liver over time \cite{2021_vaillant_transaminase} (Figure \ref{fig: Distribution of infection}).  In order to explore the changes in the  infection dynamics,   the initial conditions (equation \eqref{eq:initial}) are modified by introducing multiple points of infection.
For convenience in analyzing and comparing different scenarios, we address the solutions of the proposed model \eqref{eq:non_Healthy cells}-\eqref{eq:non_antigen}
\begin{enumerate}
\item with single point of infection as SWSPI,
\item with dual points of infection as SWDPI, 
\item with triple points of infection as SWTPI.
\end{enumerate}
\subsection{Dual-origin infection} \label{Section: Dual origin infection}
In this case, it is assumed that the infection starts from both ends ($x=0$ and $x=L$) of the liver with equal severity of infection. The initial numbers of infected hepatocytes, capsids, viruses, SVPs, and antibodies are considered to be same to those observed in single point infection, however, in this case, they are equally distributed into two endpoints. 
\subsubsection{Initial conditions}
\begin{figure}[h!]
\centering
\includegraphics[width=16.5cm, height=10.5cm]{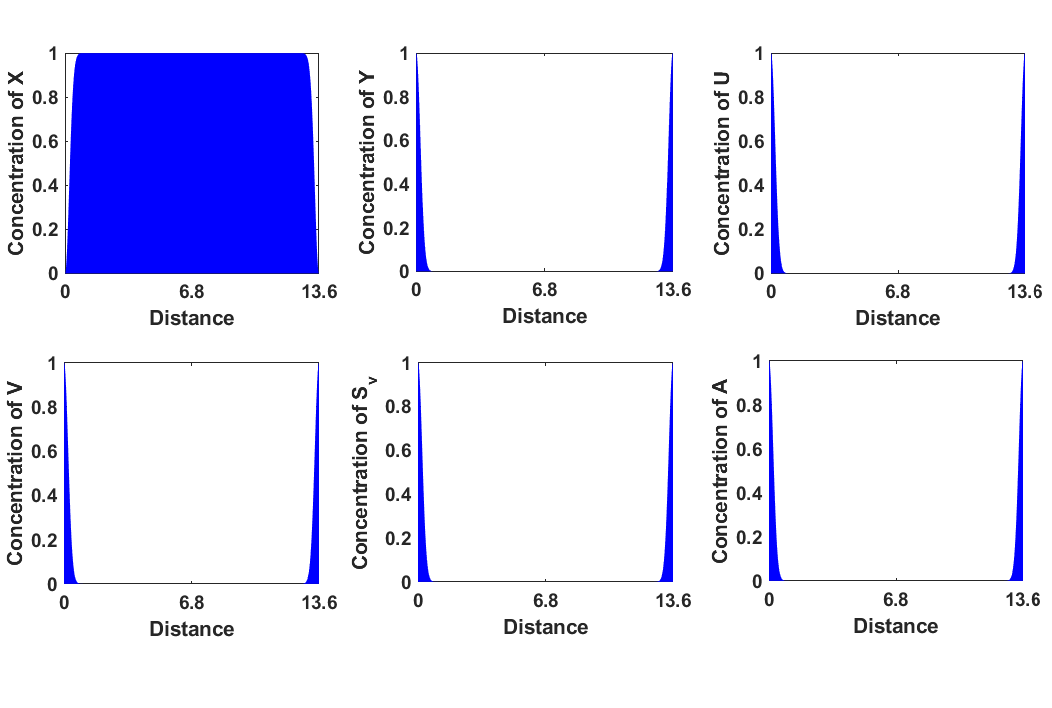}
\caption{Initial conditions: The infection initiates at $x=0$ and $x=L$ simultaneously.}
\label{Figure: Initial condition two point infection}
\end{figure}

In the context of dual-origin infection , the corresponding non-dimensional form of initial conditions are given as follows:
\begin{equation} \label{eq:initial_two_points_infection}
\left.
\begin{split}
&X(x,0)	= 1-\left\{\exp\left(-\frac{x^2}{\epsilon^*}\right)+\exp\left(-\frac{(x-1)^2}{\epsilon^*}\right)\right\},~ 0\leq x\leq 1,\\
&Y(x,0)	= \left\{\exp\left(-\frac{x^2}{\epsilon^*}\right)+\exp\left(-\frac{(x-1)^2}{\epsilon^*}\right)\right\},~0\leq x\leq 1,\\
&U(x,0)=\left\{\exp\left(-\frac{x^2}{\epsilon^*}\right)+\exp\left(-\frac{(x-1)^2}{\epsilon^*}\right)\right\},~0\leq x\leq 1,\\
&V(x,0)=\left\{\exp\left(-\frac{x^2}{\epsilon^*}\right)+\exp\left(-\frac{(x-1)^2}{\epsilon^*}\right)\right\},~0\leq x\leq 1,\\
&S_v(x,0)=\left\{\exp\left(-\frac{x^2}{\epsilon^*}\right)+\exp\left(-\frac{(x-1)^2}{\epsilon^*}\right)\right\},~0\leq x\leq 1,\\
&A(x,0)=\left\{\exp\left(-\frac{x^2}{\epsilon^*}\right)+\exp\left(-\frac{(x-1)^2}{\epsilon^*}\right)\right\},~0\leq x\leq 1.\\		 	
\end{split}
\hspace{1cm}
\right\} 
\end{equation}
In Figure \ref{Figure: Initial condition two point infection}, the initial state of each compartment is displayed for dual origin infection.
\subsection{Triple-origin infection} \label{Section: Triple origin infection}
Here, we consider that the liver is initially infected at three distinct points: both endpoints ($x=0$ and $x=L$) and the midpoint ($x=L/2$). 
Similarly, as dual-origin infection, the total number of uninfected hepatocytes,  infected hepatocytes, capsids, viruses, SVPs, and antibodies mirror those in a single-point infection.
\subsubsection{Initial conditions}
\begin{figure}[h!]
\centering
\includegraphics[width=16.5cm, height=10cm]{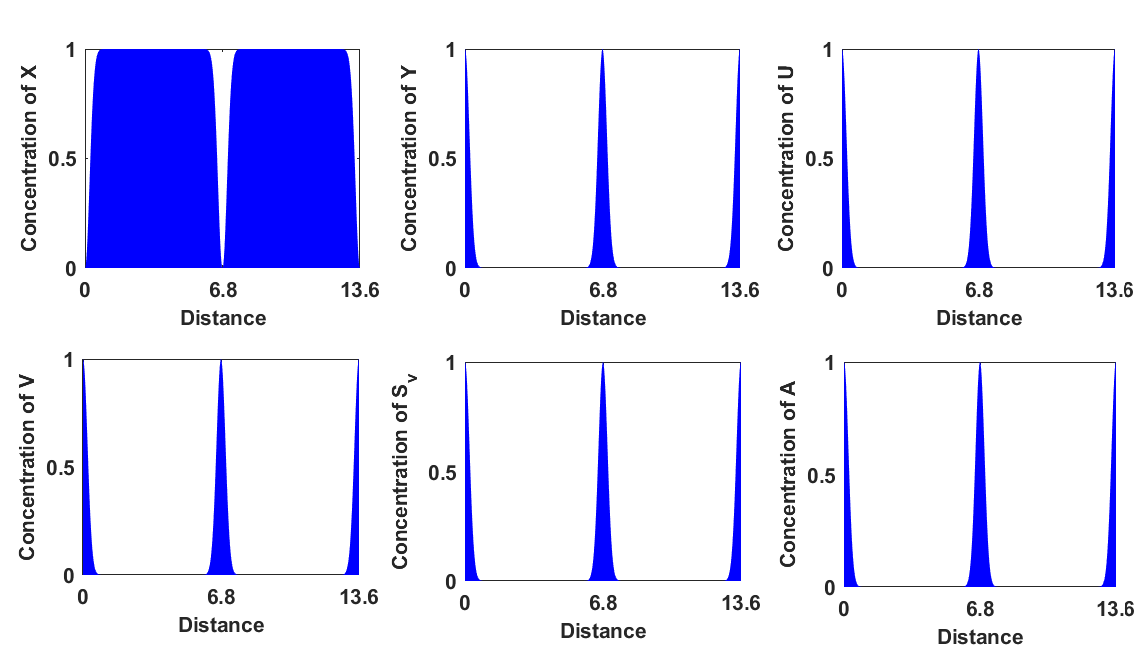}
\caption{Initial conditions: The infection initiates at $x=0$, $x=L/2$ and $x=L$ simultaneously.}
\label{Figure: Initial condition three point infection}
\end{figure}

\begin{align} \label{eq:initial_three_points_infection}
&X(x,0)	= 1-\left\{\exp\left(-\frac{x^2}{\epsilon^*}\right)+\exp\left(-\frac{\left(x-0.5\right)^2}{\epsilon^*}\right)+\exp\left(-\frac{(x-1)^2}{\epsilon^*}\right)\right\},~ 0\leq x\leq 1,\\
&Y(x,0)	= \left\{\exp\left(-\frac{x^2}{\epsilon^*}\right)+\exp\left(-\frac{\left(x-\d0.5\right)^2}{\epsilon^*}\right)+\exp\left(-\frac{(x-1)^2}{\epsilon^*}\right)\right\},~0\leq x\leq 1,\\
&U(x,0)=\left\{\exp\left(-\frac{x^2}{\epsilon^*}\right)+\exp\left(-\frac{\left(x-0.5\right)^2}{\epsilon^*}\right)+\exp\left(-\frac{(x-1)^2}{\epsilon^*}\right)\right\},~0\leq x\leq 1,\\
&V(x,0)=\left\{\exp\left(-\frac{x^2}{\epsilon^*}\right)+\exp\left(-\frac{\left(x-0.5\right)^2}{\epsilon^*}\right)+\exp\left(-\frac{(x-1)^2}{\epsilon^*}\right)\right\},~0\leq x\leq 1,\\
&S_v(x,0)=\left\{\exp\left(-\frac{x^2}{\epsilon}\right)+\exp\left(-\frac{\left(x-0.5\right)^2}{\epsilon}\right)+\exp\left(-\frac{(x-1)^2}{\epsilon}\right)\right\},~0\leq x\leq 1,\\
&A(x,0)=\left\{\exp\left(-\frac{x^2}{\epsilon^*}\right)+\exp\left(-\frac{\left(x-0.5\right)^2}{\epsilon^*}\right)+\exp\left(-\frac{(x-1)^2}{\epsilon^*}\right)\right\},~0\leq x\leq 1.\\		 	
\end{align}
Figure \ref{Figure: Initial condition three point infection} presents the initial state of each compartment for the triple-origin infection setup.
%
%
%
%
%
%
%
%
\subsubsection{Outcomes of the infection: Comparison among SWSPI, SWDPI, and SWTPI}
\begin{figure}[h!]
\centering
\includegraphics[width=17cm, height=12cm]{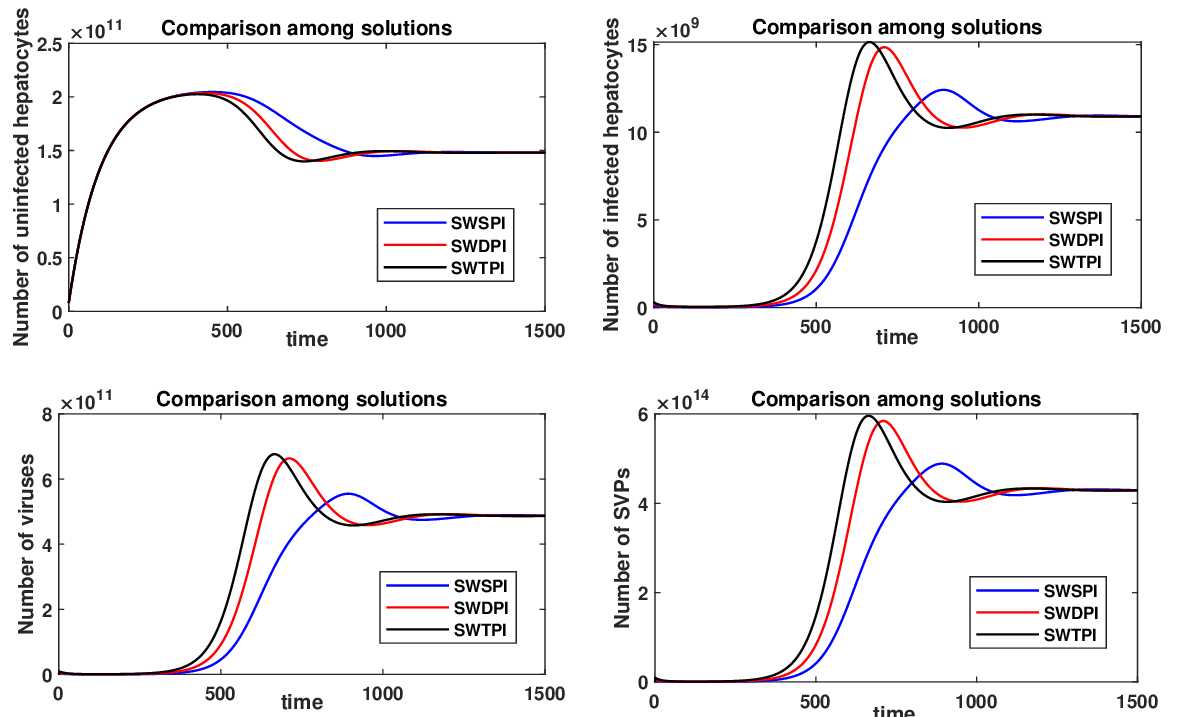}
\caption{Comparison between solutions SWSPI, SWDPI and SWTPI. SWSPI: solution with single point infection, SWDPI: solution with double points infection, SWTPI: solution with triple points infection.}
\label{Comparison_between_solution}
\end{figure}

The proposed model is solved for different initial conditions: (i) single-origin infection \eqref{eq:non_initial}, (ii) dual-origin infection \eqref{eq:initial_two_points_infection}, and (iii) triple-origin infection \eqref{eq:initial_three_points_infection}. In order to compare the model solutions,  we obtain the solution for uninfected hepatocytes, infected hepatocytes, viruses and SVPs by integrating along the spatial dimension as follows:
\begin{align}
\bar{X}(t)&=\int_{x=0}^{x=L}X(x,t)dx,\label{integration_X} \\
\bar{Y}(t)&=\int_{x=0}^{x=L}Y(x,t)dx,\label{integration_Y} \\
\bar{U}(t)&=\int_{x=0}^{x=L}U(x,t)dx,\label{integration_U} \\
\bar{V}(t)&=\int_{x=0}^{x=L}V(x,t)dx,\label{integration_V}\\
\bar{S_v}(t)&=\int_{x=0}^{x=L}S_v(x,t)dx,\label{integration_SV}\\
\bar{A}(t)&=\int_{x=0}^{x=L}A(x,t)dx,\label{integration_A} 
\end{align}
where $\bar{X}(t)$, $\bar{Y}(t)$. $\bar{V}(t)$, $\bar{S_v}(t)$ denotes the number of uninfected hepatocytes, infected hepatocytes, viruses and SVPs at time $t$, respectively. As the proposed model is solved numerically, the integrations \eqref{integration_X}, \eqref{integration_Y}, \eqref{integration_V}, \eqref{integration_SV} are converted to summation, resulting in the following expressions:
\begin{align}\label{summation}
\bar{X}(t_j)&=\sum_{i=1}^{n}X(x_n,t_j),~j=1,2,...,m,\\
\bar{Y}(t_j)&=\sum_{i=1}^{n}Y(x_n,t_j),~j=1,2,...,m,\\
\bar{U}(t_j)&=\sum_{i=1}^{n}U(x_n,t_j),~j=1,2,...,m,\\
\bar{V}(t_j)&=\sum_{i=1}^{n}V(x_n,t_j),~j=1,2,...,m,\\
\bar{S_v}(t_j)&=\sum_{i=1}^{n}S_v(x_n,t_j),~j=1,2,...,m,\\
\bar{A}(t_j)&=\sum_{i=1}^{n}A(x_n,t_j),~j=1,2,...,m,
\end{align}
where $n$ and $m$ represent the number of grid points along $x$ and $t$-axis, respectively.  In Figure \ref{Comparison_between_solution}, the model solutions for uninfected hepatocytes, infected hepatocytes, viruses and SVPs   are plotted and compared with each other.
\begin{itemize}
\item \textbf{Changes in the dynamics of uninfected hepatocytes, infected hepatocytes, viruses and SVPs}
\begin{enumerate}
\item The solutions converge to the same steady-state for all these three types of initial conditions.
\item As the number of initial points of infection increases, the peaks of infected hepatocytes, viruses and SVPs become higher.
\item If the liver is initially infected at multiple points, the number of infected cells increases rapidly for certain period of time from the time of infection.
\item For a triple-point infection, the solution peaks occur earlier compared to the peaks observed in solutions derived from dual-point or single-point infections.  
\item From a long-term perspective, the severity of the infection remains the same regardless of the number of infection points.
\end{enumerate}
\end{itemize}
%
\section{Conclusions}
\noindent
In this article, a novel HBV infection dynamics model is proposed including the impacts of sub-viral particles and antibodies. The diffusivity of each viral component (capsids, viruses, SVPs) and the HBV-specific antibodies are also incorporated into the model.   Overall, this model carry unique characteristics in the context of HBV infection. We perform non-dimensionalization on each model equation to make them easy in use. We apply forward time and centraled space (FTCS) difference scheme to solve  the proposed model numerically.
Based on the experiments, this study has yielded several key findings which are as follows:
\begin{enumerate}
\item SVPs significantly enhance intracellular viral replication and gene expression. SVPs also reduce the neutralization of virus particles by antibodies.
\item Due to the substantial presence of SVPs within the liver, the virions compartment experiences a rapid increase and  leads to a higher peak level.  The stability level of virions also increases due to the incorporation of SVPs.
\item The recycling of capsids significantly enhances the concentration of SVPs. 
\item Due to the diffusion of capsids, viruses, SVPs and antibodies, viruses spread rapidly throughout the liver. However, all model compartments ultimately converge to the same steady-state, regardless of whether diffusion is considered or not within the model.
\item As the number of initial points of infection increases, the peaks of infected hepatocytes, viruses and SVPs become higher and occur earlier.
\item From a long-term perspective, the severity of the infection remains the same regardless of the number of infection points.
\end{enumerate}
This model offers a theoretical framework for understanding the mechanism behind the exacerbation during chronic HBV infection. This model enhances our ability to accurately depict the dynamics of the infection. This model holds promise for practical application once the  clinical trials are done.
\section*{Acknowledgments}
First author would like to acknowledge the financial support obtained from CSIR (New Delhi)
under the CSIR-SRF Fellowship scheme (File No: 09/731(0171)/2019-EMR-I). 
The second author of this article thanks to Ministry of Education, Govt. of India for research fellowship and Indian Institute of Technology Guwahati, India for the support provided during the period of this work. All authors
thank the research facilities received from the Department of Mathematics, Indian Institute of Technology Guwahati, India.
\section*{Author contributions}
All authors contribute equally.
\section*{Conflict of interest}
The authors declare no potential conflict of interests.
\section*{Data Availability Statement}
Data sharing is not applicable to this article.


\begin{thebibliography}{10}
	
	\bibitem{WHO_2021}
	Hepatitis b.
	\newblock \url{https://www.who.int/news-room/fact-sheets/detail/hepatitis-b}.
	\newblock 27 July 2021.
	
	\bibitem{2008_Min}
	Lequan Min, Yongmei Su, and Yang Kuang.
	\newblock Mathematical analysis of a basic virus infection model with
	application to hbv infection.
	\newblock {\em The Rocky Mountain Journal of Mathematics}, 38(5):1573--1585,
	2008.
	
	\bibitem{2016_jun_nakabayashi}
	Jun Nakabayashi.
	\newblock The intracellular dynamics of hepatitis b virus (hbv) replication
	with reproduced virion “re-cycling”.
	\newblock {\em Journal of Theoretical Biology}, 396:154--162, 2016.
	
	\bibitem{2018_fatehi_nkcell}
	F~Fatehi Chenar, YN~Kyrychko, and KB~Blyuss.
	\newblock Mathematical model of immune response to hepatitis b.
	\newblock {\em Journal of Theoretical Biology}, 447:98--110, 2018.
	
	\bibitem{2015_Murray}
	John~M Murray and Ashish Goyal.
	\newblock In silico single cell dynamics of hepatitis b virus infection and
	clearance.
	\newblock {\em Journal of Theoretical Biology}, 366:91--102, 2015.
	
	\bibitem{2021_ciupe_early}
	Stanca~M Ciupe, Naveen~K Vaidya, and Jonathan~E Forde.
	\newblock Early events in hepatitis b infection: the role of inoculum dose.
	\newblock {\em Proceedings of the Royal Society B}, 288(1944):20202715, 2021.
	
	\bibitem{2021_prifti}
	Georgia-Myrto Prifti, Dimitrios Moianos, Erofili Giannakopoulou, Vasiliki
	Pardali, John~E Tavis, and Grigoris Zoidis.
	\newblock Recent advances in hepatitis b treatment.
	\newblock {\em Pharmaceuticals}, 14(5):417, 2021.
	
	\bibitem{2005_gripon_efficient}
	Philippe Gripon, Isabelle Cannie, and Stephan Urban.
	\newblock Efficient inhibition of hepatitis b virus infection by acylated
	peptides derived from the large viral surface protein.
	\newblock {\em Journal of virology}, 79(3):1613--1622, 2005.
	
	\bibitem{2013_gerlich_medical}
	Wolfram~H Gerlich.
	\newblock Medical virology of hepatitis b: how it began and where we are now.
	\newblock {\em Virology journal}, 10(1):1--25, 2013.
	
	\bibitem{2015_luckenbaugh_genome}
	L~Luckenbaugh, KM~Kitrinos, WE~Delaney~IV, and J~Hu.
	\newblock Genome-free hepatitis b virion levels in patient sera as a potential
	marker to monitor response to antiviral therapy.
	\newblock {\em Journal of viral hepatitis}, 22(6):561--570, 2015.
	
	\bibitem{2017_hu_complete}
	Jianming Hu and Kuancheng Liu.
	\newblock Complete and incomplete hepatitis b virus particles: formation,
	function, and application.
	\newblock {\em Viruses}, 9(3):56, 2017.
	
	\bibitem{2009_garcia_drastic}
	Tamako Garcia, Jisu Li, Camille Sureau, Kiyoaki Ito, Yanli Qin, Jack Wands, and
	Shuping Tong.
	\newblock Drastic reduction in the production of subviral particles does not
	impair hepatitis b virus virion secretion.
	\newblock {\em Journal of virology}, 83(21):11152--11165, 2009.
	
	\bibitem{1996_Nowak}
	Martin~A Nowak, Sebastian Bonhoeffer, Andrew~M Hill, Richard Boehme, Howard~C
	Thomas, and Hugh McDade.
	\newblock Viral dynamics in hepatitis b virus infection.
	\newblock {\em Proceedings of the National Academy of Sciences},
	93(9):4398--4402, 1996.
	
	\bibitem{2005_murray_dynamics}
	John~M Murray, Stefan~F Wieland, Robert~H Purcell, and Francis~V Chisari.
	\newblock Dynamics of hepatitis b virus clearance in chimpanzees.
	\newblock {\em Proceedings of the National Academy of Sciences},
	102(49):17780--17785, 2005.
	
	\bibitem{2007_Ciupe_Role}
	Stanca~M Ciupe, Ruy~M Ribeiro, Patrick~W Nelson, Geoffrey Dusheiko, and Alan~S
	Perelson.
	\newblock The role of cells refractory to productive infection in acute
	hepatitis b viral dynamics.
	\newblock {\em Proceedings of the National Academy of Sciences},
	104(12):5050--5055, 2007.
	
	\bibitem{2007_Dahari}
	Harel Dahari, Arthur Lo, Ruy~M Ribeiro, and Alan~S Perelson.
	\newblock Modeling hepatitis c virus dynamics: Liver regeneration and critical
	drug efficacy.
	\newblock {\em Journal of Theoretical Biology}, 247(2):371--381, 2007.
	
	\bibitem{2007_wang_wang}
	Kaifa Wang and Wendi Wang.
	\newblock Propagation of hbv with spatial dependence.
	\newblock {\em Mathematical Biosciences}, 210(1):78--95, 2007.
	
	\bibitem{2009_Eikenberry}
	Steffen Eikenberry, Sarah Hews, John~D Nagy, and Yang Kuang.
	\newblock The dynamics of a delay model of hbv infection with logistic
	hepatocyte growth.
	\newblock {\em Mathematical Biosciences and Engineering}, 6:1--17, 2009.
	
	\bibitem{2010_Hews}
	Sarah Hews, Steffen Eikenberry, John~D Nagy, and Yang Kuang.
	\newblock Rich dynamics of a hepatitis b viral infection model with logistic
	hepatocyte growth.
	\newblock {\em Journal of Mathematical Biology}, 60(4):573--590, 2010.
	
	\bibitem{2011_Jun_nakabayashi}
	Jun Nakabayashi and Akira Sasaki.
	\newblock A mathematical model of the intracellular replication and within host
	evolution of hepatitis type b virus: Understanding the long time course of
	chronic hepatitis.
	\newblock {\em Journal of Theoretical Biology}, 269(1):318--329, 2011.
	
	\bibitem{2018_Danane_mathematical}
	Jaouad Danane and Karam Allali.
	\newblock Mathematical analysis and treatment for a delayed hepatitis b viral
	infection model with the adaptive immune response and dna-containing capsids.
	\newblock {\em High-throughput}, 7(4):35, 2018.
	
	\bibitem{2018_Guo}
	Ting Guo, Haihong Liu, Chenglin Xu, and Fang Yan.
	\newblock Global stability of a diffusive and delayed hbv infection model with
	hbv dna-containing capsids and general incidence rate.
	\newblock {\em Discrete \& Continuous Dynamical Systems-B}, 23(10):4223, 2018.
	
	\bibitem{2021_Fatehi}
	Farzad Fatehi, Richard~J Bingham, Eric~C Dykeman, Nikesh Patel, Peter~G
	Stockley, and Reidun Twarock.
	\newblock An intracellular model of hepatitis b viral infection: An in silico
	platform for comparing therapeutic strategies.
	\newblock {\em Viruses}, 13(1):11, 2021.
	
	\bibitem{2023_sutradhar_fractional}
	Rupchand Sutradhar and D.~C. Dalal.
	\newblock Fractional-order models of hepatitis b virus infection with recycling
	effects of capsids.
	\newblock {\em Mathematical Methods in the Applied Sciences},
	46(14):15599--15625.
	
	\bibitem{2019_cao_cryo}
	Jianhao Cao, Junchang Zhang, Yanmeng Lu, Shuhong Luo, Jingqiang Zhang, and Ping
	Zhu.
	\newblock Cryo-em structure of native spherical subviral particles isolated
	from hbv carriers.
	\newblock {\em Virus research}, 259:90--96, 2019.
	
	\bibitem{2017_rydell_hepatitis}
	Gustaf~E Rydell, Kasthuri Prakash, Hel{\'e}ne Norder, and Magnus Lindh.
	\newblock Hepatitis b surface antigen on subviral particles reduces the
	neutralizing effect of anti-hbs antibodies on hepatitis b viral particles in
	vitro.
	\newblock {\em Virology}, 509:67--70, 2017.
	
	\bibitem{2008_chai_properties}
	Ning Chai, Ho~Eun Chang, Emmanuelle Nicolas, Ziying Han, Michal Jarnik, and
	John Taylor.
	\newblock Properties of subviral particles of hepatitis b virus.
	\newblock {\em Journal of virology}, 82(16):7812--7817, 2008.
	
	\bibitem{1998_bruns_enhancement}
	Michael Bruns, Stefan Miska, Sylvie Chassot, and Hans Will.
	\newblock Enhancement of hepatitis b virus infection by noninfectious subviral
	particles.
	\newblock {\em Journal of virology}, 72(2):1462--1468, 1998.
	
	\bibitem{1993_klingmuller_hepadnavirus}
	U~Klingm{\"u}ller and H~Schaller.
	\newblock Hepadnavirus infection requires interaction between the viral pre-s
	domain and a specific hepatocellular receptor.
	\newblock {\em Journal of Virology}, 67(12):7414--7422, 1993.
	
	\bibitem{2021_wu_dynamical}
	Peng Wu and Hongyong Zhao.
	\newblock Dynamical analysis of a nonlocal delayed and diffusive hiv latent
	infection model with spatial heterogeneity.
	\newblock {\em Journal of the Franklin Institute}, 358(10):5552--5587, 2021.
	
	\bibitem{2023_wu_spatial}
	Peng Wu, Xiunan Wang, and Zhaosheng Feng.
	\newblock Spatial and temporal dynamics of sars-cov-2: Modeling, analysis and
	simulation.
	\newblock {\em Applied mathematical modelling}, 113:220--240, 2023.
	
	\bibitem{2023_miao_dynamics}
	Hui Miao, Meiyan Jiao, et~al.
	\newblock Dynamics of a diffusive hbv infection model with capsids, two delays,
	and cell-to-cell transmissions.
	\newblock {\em Discrete Dynamics in Nature and Society}, 2023, 2023.
	
	\bibitem{2009_xu_hbv}
	Rui Xu and Zhien Ma.
	\newblock An hbv model with diffusion and time delay.
	\newblock {\em Journal of Theoretical Biology}, 257(3):499--509, 2009.
	
	\bibitem{2008_wang_dynamics}
	Kaifa Wang, Wendi Wang, and Shiping Song.
	\newblock Dynamics of an hbv model with diffusion and delay.
	\newblock {\em Journal of Theoretical Biology}, 253(1):36--44, 2008.
	
	\bibitem{2023_sutradhar_recycling}
	Rupchand Sutradhar and D.~C. Dalal.
	\newblock Re-cycling of dna-containing capsids enhances hepatitis b.
	
	\bibitem{2012_prange_host}
	Reinhild Prange.
	\newblock Host factors involved in hepatitis b virus maturation, assembly, and
	egress.
	\newblock {\em Medical microbiology and immunology}, 201(4):449--461, 2012.
	
	\bibitem{2014_ciupe_antibody}
	Stanca~M Ciupe, Ruy~M Ribeiro, and Alan~S Perelson.
	\newblock Antibody responses during hepatitis b viral infection.
	\newblock {\em PLoS computational biology}, 10(7):e1003730, 2014.
	
	\bibitem{2018_chenar_mathematical}
	F~Fatehi Chenar, YN~Kyrychko, and KB~Blyuss.
	\newblock Mathematical model of immune response to hepatitis b.
	\newblock {\em Journal of theoretical biology}, 447:98--110, 2018.
	
	\bibitem{2019_huang_mathematical}
	Kuo-Sheng Huang, Yu-Chiau Shyu, Chih-Lang Lin, and Feng-Bin Wang.
	\newblock Mathematical analysis of an hbv model with antibody and spatial
	heterogeneity.
	\newblock {\em Math. Biosci. Eng}, 17(2):1820--1837, 2019.
	
	\bibitem{chaplain2006mathematical}
	Mark~AJ Chaplain and G~Lolas.
	\newblock Mathematical modelling of cancer invasion of tissue: dynamic
	heterogeneity.
	\newblock {\em Networks Heterog. Media}, 1(3):399--439, 2006.
	
	\bibitem{2019_Liver_size}
	Kamrul Islam.
	\newblock Role of ultrasonography to detect liver size.
	\newblock {\em Ultrasound in Medicine and Biology}, 45:S96, 2019.
	
	\bibitem{2006_Murray}
	John~M Murray, Robert~H Purcell, and Stefan~F Wieland.
	\newblock The half-life of hepatitis b virions.
	\newblock {\em Hepatology}, 44(5):1117--1121, 2006.
	
	\bibitem{2017_meskaf_optimal}
	Adil Meskaf, Karam Allali, and Youssef Tabit.
	\newblock Optimal control of a delayed hepatitis b viral infection model with
	cytotoxic t-lymphocyte and antibody responses.
	\newblock {\em International Journal of Dynamics and Control}, 5:893--902,
	2017.
	
	\bibitem{2020_vaillant_hbsag}
	Andrew Vaillant.
	\newblock Hbsag, subviral particles, and their clearance in establishing a
	functional cure of chronic hepatitis b virus infection.
	\newblock {\em ACS infectious diseases}, 7(6):1351--1368, 2020.
	
	\bibitem{2021_vaillant_transaminase}
	Andrew Vaillant.
	\newblock Transaminase elevations during treatment of chronic hepatitis b
	infection: Safety considerations and role in achieving functional cure.
	\newblock {\em Viruses}, 13(5):745, 2021.
	
\end{thebibliography}
\end{document}